%% file: ms.tex
\documentclass[]{emulateapj}

\usepackage{amssymb}

\begin{document}

\title{Quasar Broad Absorption Line Variability on Multi-Year Time Scales}

\author{Robert R. Gibson, W. N. Brandt, Donald P. Schneider}
\affil{Department of Astronomy and Astrophysics\\Pennsylvania State University\\525 Davey Laboratory\\University Park, PA, 16802}
\email{rgibson@astro.psu.edu}

\and
\author{S. C. Gallagher}
\affil{Department of Physics and Astronomy\\University of California -- Los Angeles\\Los Angeles, CA, 90095}
\affil{Physics and Astronomy Department\\University of Western Ontario\\London, ON N6A 3K7, CANADA}

\shorttitle{BAL Variation on Multi-Year Time Scales}
\shortauthors{Gibson et al.}

\journalinfo{Accepted to ApJ. (c) Copyright 2008.  The American Astronomical Society.  All rights reserved.  Printed in U.S.A.}

\clearpage

\begin{abstract}
We use quantitative metrics to characterize the variation of \ion{C}{4} $\lambda 1549$ broad absorption lines (BALs) over 3--6 (rest-frame) years in a sample of 13 quasars at $1.7 \le z \le 2.8$ and compare the results to previous studies of BAL variability on shorter time scales.  The strong BALs in our study change in complex ways over 3--6 yr.  Variation occurs in discrete regions which are only a few thousand km~s$^{-1}$ wide, and the distribution of the change in absorption equivalent width broadens over time.  We constrain the typical \ion{C}{4} BAL lifetime to be at least a few decades.  While we do not find evidence to support a scenario in which the variation is primarily driven by photoionization on multi-year time scales, there is some indication that the variation is produced by changes in outflow geometry.  We do not observe significant changes in the BAL onset velocity, indicating that the absorber is either far from the source or is being continually replenished and is azimuthally symmetric.

It is not possible in a human lifetime to expand the time scales in our study by more than a factor of a few using optical spectroscopy.  However, the strong variation we have observed in some BALs indicates that future studies of large numbers of BAL QSOs will be valuable to constrain BAL lifetimes and the physics of variation.

\end{abstract}


\section{INTRODUCTION\label{introSec}}

Broad Absorption Line Quasars (BAL QSOs) exhibit broad, ultraviolet (UV) line absorption troughs spanning thousands, or tens of thousands, of km~s$^{-1}$.  By studying the evolution of these dramatic features, we hope to learn about the structure and dynamics of absorbing material in active galactic nuclei (AGNs).  Because QSO redshifts of $z \sim 2$ are needed to shift high-ionization BAL lines into the optical waveband, observations of BAL QSOs must span decades in order to characterize multi-year BAL variability for high-ionization lines such as \ion{C}{4} $\lambda$1549.

BAL troughs are present in about 20\% of QSOs \citep[e.g.,][]{hf03}, produced by absorption from lines such as \ion{C}{4} $\lambda$1549 and \ion{Mg}{2} $\lambda$2799.  BAL absorption can extend to very high velocities, although above $\approx0.1 c$ absorption from \ion{C}{4} $\lambda 1549$ becomes confused with lines from \ion{Si}{4} and \ion{O}{4}.  BAL QSOs with broad absorption from low ionization stages such as \ion{Mg}{2} or \ion{Al}{3} are classified as ``low-ionization BAL'' (LoBAL) QSOs, while those with absorption only from ions at higher ionization stages are called ``high-ionization BAL'' (HiBAL) QSOs.

It is commonly thought that BAL outflows are seen when QSOs are observed at large inclination angles with respect to accretion disk.  In this model, the line of sight to the central source passes through an equatorial disk wind, which imprints a broad absorption signature on the continuum.  BAL outflows would be present in every QSO, but only cover about 20\% of the QSO sky.  Modeling this scenario has successfully reproduced many properties of BAL absorption \citep[e.g.,][]{mcgv95, psk00}.  Other studies have modeled the BAL outflow as an orientation-independent evolutionary effect, caused by the expulsion of a thick shroud of gas and dust \citep[e.g.,][]{vwk93, gbd06} which may have been deposited in the nucleus by galaxy collisions \citep[e.g.,][]{cs01, jsbbchl01, gbwrcf02}.

BAL outflows have a complex structure.  The covering factor, or geometric fraction of the emission which is obscured by the absorber(s), can vary strongly across the absorption trough as a function of outflow velocity \citep[e.g.,][]{dabgwlpk01}.  As a result, the absorption is often saturated but not black, so that the outflow geometry (with respect to the emitter) determines the absorption profile \citep[e.g.,][]{ablgwbd99}.  The ionization level also apparently varies with velocity, and, at least in the case of PG 0946+301, the covering factor increases with the ionization level of the absorber \citep{akdjb99}.  Spectropolarimetric studies have found that the emission at the bottom of deep troughs is highly polarized and likely scattered around the absorber \citep[e.g,][]{ocmtgm99}.

\begin{figure} [ht]
  \begin{center}
      \includegraphics[width=2.3in, angle=270]{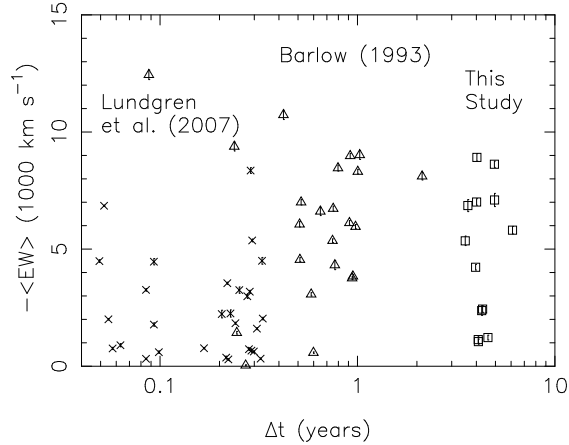}
      \caption{\label{avgEWVsTFig}A comparison of the rest-frame EWs (averaged over two epochs) of the \ion{C}{4} BALs for the sources in B93 (triangles), L07 (crosses), and our study (squares).  For each source, $-\langle EW \rangle$ is plotted against the rest-frame time between epochs, $\Delta t_{sys}$.  Vertical bars mark the error in EW.  Note that the $x$-axis is logarithmic.}
   \end{center}
\end{figure}

In this study, we take advantage of the overlap between the Large Bright Quasar Survey \citep[LBQS;][and references therein]{hfc95} and the Sloan Digital Sky Survey \citep[SDSS;][]{y+00} to examine variation in a sample of 13 BAL QSOs over rest-frame time scales of 3--6 yr.\footnote{Throughout this work, all durations are given in the rest frame defined by the QSO redshift, unless otherwise noted.}  Previous studies have focused on individual sources (\S\ref{indivBALVarSec}) or on a larger number of sources over shorter ($< 1$~yr) time scales (\S\ref{bALVarSampleSec}).  For comparison, we plot BAL equivalent widths (EWs) and times between observations in Figure~\ref{avgEWVsTFig} for our study and the samples of \citet[][hereafter B93] {b93} and \citet[][hereafter L07]{lwbhsyvb07}.\footnote{The method of calculating EWs varies slightly among the studies, but the differences should not strongly affect comparisons between the samples.}  In this paper, we adopt the convention that absorption features have negative EWs, and the EWs refer to rest-frame values.

BAL variation may be driven by several processes working on different time scales.  For sufficiently dense absorbers, photoionization time scales can be weeks or less \citep[e.g.,][]{kk01}.  Flow time scales for structural variation can be much longer, and proceed independently of the continuum.  For example, at a speed of 10,000~km~s$^{-1}$, it would take $\approx$3~yr to cross a region of size $\sim10^{17}$~cm ($\sim10,000$ gravitational radii for a $10^8 M_{\sun}$ black hole).  Eventually, surveys of large numbers of BALs should be able to distinguish variation trends at different time scales.  This study takes one step toward that goal by characterizing multi-year variation in a sample of \ion{C}{4} BALs in the context of previous studies on shorter time scales.

In the remainder of this introduction, we briefly review studies of BAL variation for individual BAL sources (\S\ref{indivBALVarSec}); for narrower, ``mini-BAL'' absorbers (\S\ref{narrowerAbsVarSec}); and also for samples of objects with BAL absorbers (\S\ref{bALVarSampleSec}).  Throughout this paper, we use a cosmology with $H_0 = 72$, $\Omega_{M} = 0.3$, and $\Omega_{\Lambda} = 0.7$.

\subsection{BAL Variation in Individual Objects\label{indivBALVarSec}}

In this section, we summarize recent observations of BAL variability in {\it individual} sources to give context for our study.  High-velocity absorption components have varied over a few months to a few years in some sources \citep{fwmt87, sp88, tgfw88, vi01}.  In other cases, the variation has occurred across a wide range of velocities and ionization stages \citep[e.g.,][]{a+01} or in multiple absorption troughs \citep[e.g.,][]{h+02}.  Dramatic X-ray absorption variation has been observed in PG 1004+130 \citep{mbglwgs06} and PG 2112+059 \citep{gbwccl04}, while UV absorption also varied subtly in the latter source.

Some QSOs have varied so strongly that the BAL status of the object has changed.  The \ion{Mg}{2} absorber of LBQS 0103--2753A disappeared over 6 yr, changing the technical classification of that source from LoBAL to HiBAL \citep{jsbbchl01}.  Narrow \ion{C}{4} and \ion{Si}{4} absorption lines in the radio-loud QSO TEX 1726+344 broadened over 3.5 yr, changing its classification from a non-BAL to a BAL QSO \citep{m02}.

Photoionization has been proposed as a driver of BAL variation \citep[e.g.,][]{bjb89}, but this explanation can be problematic.  In several cases, absorber variation does not appear to track the continuum \citep{bjbwmk92, mon96}.  The absorber may be too tenuous to change its ionization state on continuum variation time scales, or BAL variation may be governed by additional structural factors.

Despite the high velocities and complex structures commonly observed in BALs, there have only been a few claims for detections of acceleration in BAL troughs.  \citet{vi01} presented a spectrum of Q1303+308 which showed strong BAL variation over 6 yr, particularly in the highest-velocity components.  They also reported that the outflow velocity had increased in magnitude by 55~km~s$^{-1}$, corresponding to an acceleration of $\approx$0.03~cm~s$^{-2}$.  \citet{rvs02} point out that one high-velocity component of the \ion{Na}{1} D line of the nearby QSO Mrk 231 accelerated over about 10 yr.  Recently, \citet{hsher07} compared SDSS and ESO Very Large Telescope (VLT) spectra of SDSS~J024221.87+004912.6, finding evidence for an average acceleration of the \ion{C}{4} BAL trough of 0.15~cm~s$^{-2}$ over 1.4~yr.

\subsection{Variation in Narrower Absorption Systems\label{narrowerAbsVarSec}}

Narrower intrinsic absorption lines share many properties in common with BALs \citep[e.g.,][]{akdjb99}, so it is worthwhile also to consider the variation of mini-BALs (i.e., broad absorption troughs which are somewhat narrower than the 2000~km s$^{-1}$ required for a BAL).  Mrk~231 shows broad ($\approx 1500$~km~s$^{-1}$ wide) absorption lines from several different ions \citep{bcafps77} and has been the subject of many investigations.  A new velocity component for \ion{Na}{1} and \ion{He}{1} appeared between 1984 December and 1988 May \citep{bmmp91}.  \ion{Na}{1} D line components have continued to vary over the last decade, and the optical polarization of this source has also been observed to vary \citep[e.g.,][]{gssbchhb05}.

Recently, \citet{meck07} reported monitoring of the QSO HS 1603+3820 ($z = 2.54$) which hosts a mini-BAL.  They tracked substructure variation in the mini-BAL trough over about 1.2~yr at high spectral resolution with the {\it Subaru} High Dispersion Spectrograph (HDS).  The EWs of individual components of the mini-BAL varied together, and this variation was correlated with changes in the covering factor of the absorber.

\subsection{Studies of BAL Variability in Samples\label{bALVarSampleSec}}

In an early study of BAL variability, B93 monitored spectra of 23 BAL QSOs over times up to 1~yr.  BALs in four QSOs changed dramatically during the study, while those in 11 more objects changed at least marginally.  The BAL features varied in intensity rather than in velocity (within the limits of the data), and there was some tendency for BAL changes to correlate with broad-band continuum variation.  These findings supported the hypothesis that photoionization is the primary driver of BAL variability.

Recently, L07 searched for variability in the spectra of 29 BAL QSOs which had been observed by the SDSS in two epochs separated by up to 4 months.  They allowed a somewhat looser definition of ``BAL,'' retaining in their sample objects with absorption velocity widths down to 1000~km~s$^{-1}$.  They found that the magnitude of the fractional change in EW, $|\Delta EW / \langle EW \rangle|$, was greatest over the longest times between epochs and for the weakest BALs.  The strongest fractional changes in EW occurred at outflow velocities of --12,000 to --16,000~km~s$^{-1}$.  While half of the varying absorption troughs showed changes in velocity width, variation in trough depth was more common.  No correlation was found between changes in the BAL EW and UV continuum variability.

\section{BAL SAMPLE SELECTION\label{balSampleSelectionSec}}

The LBQS contains spectra of 1055 QSOs over a redshift range $0.2 \le z \le 3.4$ obtained between 1986 and 1989 \citep{hfc95}.  The resolution of the spectra used in the current study is $\approx$6~\AA\ (FWHM), with a sampling of $\approx$2.5~\AA\ per bin (in the observed frame) \citep{mwahffcm91}.  The LBQS spectra were kindly provided by P. Hewett for this study.  BAL QSOs in the LBQS have been cataloged by \citet{hf03}; we refer to these objects as ``HF03 BAL QSOs.''

The SDSS QSO spectra have been obtained since the year 2000 at a spectral resolution of $\approx$3~\AA\ with $\approx$1~\AA\ per spectral bin (in the observed frame).  The SDSS QSO catalog for Data Release 5 (DR5) identifies 77,429 QSOs in SDSS DR5 and gives spectroscopic redshifts for these sources \citep{s+07}.  We have searched in the public SDSS database for objects that correspond to LBQS QSOs in order to obtain the maximal set of sources which have both LBQS and SDSS spectroscopy available.

Most LBQS spectra were obtained using the Multiple Mirror Telescope (MMT) spectrograph, with spectral coverage from $3200 \la \lambda \la 6400$~\AA.  The LBQS spectra extend out to $\approx7500$~\AA, but are contaminated by the second-order spectrum for $\lambda \ga 6500$~\AA, and we discard this wavelength range.  The SDSS spectra cover longer wavelengths, $3800 \la \lambda \la 9200$~\AA.  QSO redshifts of $1.6 \la z \la 3.1$ are required to fit continua and cover the \ion{C}{4} absorption region from 1400--1550~\AA\ in both spectra.  We converted the LBQS air wavelengths to vacuum wavelengths for comparison with SDSS spectra \citep{m91}.

There are 127 QSOs (with or without BALs) which have full \ion{C}{4} region coverage in both surveys.  No objects have coverage of both the \ion{C}{4} and \ion{Mg}{2} regions in both epochs, so we are unable to study \ion{C}{4} BALs in relation to \ion{Mg}{2} absorption (or lack of absorption).  We have considered analyzing \ion{Mg}{2} BALs individually, but several factors complicate tests for broad \ion{Mg}{2} absorption in our spectra.  Broad emission features in the \ion{Mg}{2} region obscure absorption and the underlying continuum, and the spectra do not cover the red side of the emission region in many cases.

We have included one BAL QSO, LBQS 0021--0213, which we previously observed with the Hobby-Eberly Telescope ({\it HET}) Low-Resolution Spectrograph (LRS) on 2001 December 10 at a sufficient resolution to include in our study.  (This source was not observed by the SDSS.)  The source was observed with a 2\arcsec~slit using the g2 grism and the GG385 filter, giving a resolution of 8~\AA.  We reduced the spectrum using standard procedures with the Image Reduction and Analysis Facility (IRAF) version 2.12.  It was flux-calibrated using observations of a standard star taken the same night.  While the absolute flux measurement for this source is not known with sufficient accuracy to be used in our study, the structure of broad absorption features (with a continuum divided out) is reliably represented in the spectrum.  We converted air wavelengths to vacuum wavelengths as we did for LBQS spectra.

The error in the LBQS flux calibration is estimated to be $\approx$15\% \citep{hfcfwmam91}.  We re-normalize the SDSS spectra so that the synthetic filter fluxes match the photometric fluxes.  We divide out the continuum in most of our study, rendering the absolute flux normalization unimportant.  In cases where we do use the continuum flux, we assume the SDSS error is small compared to the LBQS error.

\subsection{Continuum Fitting\label{contFitSec}}

In this section, we describe our method of fitting continua and emission line profiles to LBQS, SDSS, and {\it HET} spectra of all 128 sources with two epochs of complete coverage in the \ion{C}{4} region.  We correct all spectra for Galactic extinction using the reddening curve of \citet{ccm89} with the near-UV extension of \citet{o94}.  We obtain $E(B-V)$ from the NASA Extragalactic Database (NED)\footnote{http://nedwww.ipac.caltech.edu/}, which uses the dust maps of \citet{sfd98}.

We convert all spectra to the rest frame using the SDSS DR3 QSO catalog redshift when available, otherwise we use the redshift assigned by the SDSS data-processing pipeline.\footnote{The DR5 QSO catalog \citep{s+07} had not been published when we began our study, but the only source with a redshift different from that of the DR5 catalog is 002444.11+003221.4 (at $z = 0.40$) which differs by $|\Delta z| = 0.007$.  This source does not affect our study.}  (These redshifts also agree with the redshifts assigned by the LBQS.)  Some of the emission lines used by the SDSS pipeline to determine redshifts may be distorted by BAL features, but it does not appear that this effect has significantly impacted SDSS redshift determination for our sources.  The emission from \ion{C}{3}] $\lambda 1909$ is only weakly absorbed, and our fits of this line in the SDSS spectra are not greatly influenced by \ion{Al}{3} $\lambda 1857$ emission 40~\AA\ to the blue side.  Comparing the redshift of the \ion{C}{3}] emission line in our fits to the SDSS systemic redshifts, we find a mean difference $\langle\Delta z\rangle \equiv \langle z_{SDSS} - z_{CIII]} \rangle = 0.007$ with a maximum difference $\max(|\Delta z|) = 0.014$.  An RMS difference of 0.006 in redshift determination has been observed in sources observed multiple times in the SDSS \citep{wvkspbrb05, s+07}.  Therefore, we do not believe that BAL absorption has led to large errors in redshift for our sample.

We fit a continuum to the following regions (when present), which are expected to be relatively free of line emission and absorption features:  1250--1350, 1600--1800, 1950--2050, 2150--2250, and 2950--3700~\AA\ (rest frame).  At each iteration, we ignore any spectral bins which deviate by more than 3$\sigma$ from the previous iteration's model fit.  This allows us to exclude additional regions which contain unpredictable broad emission or absorption structures such as those seen in some cases on the red side of \ion{C}{4}~$\lambda 1549$ and the blue side of \ion{Mg}{2}~$\lambda$2799.  We fit a power law continuum intrinsically reddened using the Small Magellanic Cloud (SMC) curve of \citet{p92}.  As in \citet{hshrcsvjbs04}, we find that SMC-like reddening reproduces the observed spectra more accurately than a Galactic reddening model does.  Although the underlying continuum structure may be more complex than a reddened power law (due, e.g., to breaks in the continuum), our experience with the sources in this paper (and also a large number of SDSS spectra not used in this study) indicate that a reddened power law reproduces the continuum structure well with a small number of fit parameters.

We use our best-fit continuum model to calculate $F_{\nu}(2500~$\AA$)$, the continuum flux density at (rest-frame) 2500~\AA.  Because broad emission line features can contaminate the spectrum at 2500~\AA, we prefer to use the continuum fit rather than a direct measurement of the continuum at that wavelength.  This method also allows extrapolation of the continuum to 2500~\AA\ when that region is not present in the observed spectrum.  We similarly calculate $F_{\nu}(1400~$\AA$)$, the continuum flux density at (rest-frame) 1400~\AA.  For the BAL QSOs in our study, the continuum is reasonably well-constrained at 1400~\AA.

With the continuum fixed, we fit Voigt emission line profiles, corresponding to \ion{Si}{4} $\lambda1399$, \ion{C}{4} $\lambda1549$, \ion{C}{3}] $\lambda1909$, and \ion{Mg}{2} $\lambda2799$, when these wavelengths are present in our spectrum.  We use a single Voigt profile for each emission line, as additional emission structure is not generally required by our data.  We fit each emission line iteratively.  At each iteration we ignore ``absorbed'' bins which are more than 2.5$\sigma$ below the previous model fit.  We do not consider the Voigt profiles to be physical models.  They represent the overall emission line profile well with a small number of fit parameters.

\begin{figure} [ht]
  \begin{center}
      \includegraphics[width=2in, angle=270]{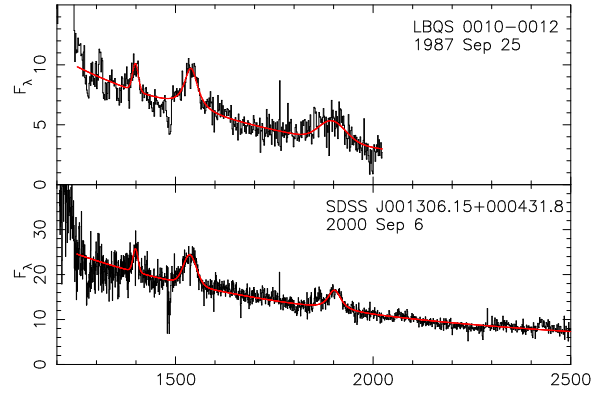}
      \caption{\label{contFitsFig0}Plot of the LBQS (top) and SDSS (bottom) spectra of each source.  The best model fit for the continuum and strong emission lines in each epoch is plotted with a thick red line.  The data are binned by a factor of of 2 over the instrumental sampling.  The $y$-axis is the observed-frame flux density in units of 10$^{-17}$~erg~s$^{-1}$~cm$^{-2}$~\AA$^{-1}$ at the rest-frame wavelength given by the $x$-axis.  In cases where the continuum in the \ion{C}{4} absorption region is not well-constrained, we have used the LBQS fit (with data extending to shorter wavelengths) as a guide.  See \S\ref{contFitSec} for additional details.  The sources shown and their dates of observation are indicated in the plots.  Figures for the remaining sources are available in the electronic edition of the Journal.  The printed edition contains only a sample.}
   \end{center}
\end{figure}

Visual inspection indicates that our fitting algorithm generally does an excellent job of reconstructing the underlying continuum and emission line profiles.  In a small number of cases, the continuum fits are unsatisfactory.  We manually adjust the continuum in these cases and then re-fit the emission lines as before to obtain a visually satisfactory fit.  For a few BAL QSOs where the SDSS spectrum does not extend blueward of the \ion{C}{4} absorption region, we have used the LBQS spectrum as a guide to estimate the SDSS continuum.  The final continuum fits are shown in Figure~\ref{contFitsFig0}.

Although we fit a reddened power law model to the continuum, we do not interpret our fit results as physically meaningful models.  BAL QSOs are more intrinsically reddened than non-BAL QSOs \citep{wmfh91, btbglw01, rrhsvfykb03}, and it can be difficult to determine the extent of continuum reddening when the shorter-wavelength spectral regions are strongly affected by both broad line emission and absorption features.  Differences in the telescopes and calibration used for the LBQS and SDSS further complicate comparisons between the spectra \citep[e.g., \S2.3 of ][]{hfcfwmam91}.  Because the LBQS and SDSS spectra span different wavelength ranges, we also expect that the distribution of fit parameters may differ.  The SDSS spectra cover a broader range of longer wavelengths, providing tighter constraints on the power law continuum and weaker constraints on the intrinsic reddening.

Throughout this study, we refer to the ratio spectrum, $R(\lambda)$, which is constructed by dividing the observed spectral flux density by the continuum and emission line fit model.  For the \ion{C}{4} region (1400--1550~\AA), this includes the \ion{C}{4} and \ion{Si}{4} emission lines.  In velocity space, we call the ratio spectrum $R(v)$.  We consider outflow velocities and absorption equivalent widths to be negative throughout.  Changes in measured quantities are calculated by subtracting the value in the earlier epoch from that of the later epoch.  With these conventions, $\Delta EW < 0$ corresponds to an increase in absorption strength over time.

\subsection{Identifying BAL QSOs\label{identifyBALsSec}}

BALs are traditionally identified according to the balnicity index, $BI$, which approximately expresses the absorption EW in km~s$^{-1}$ of a trough which spans at least 2000~km~s$^{-1}$ and is offset by a ``detachment velocity'' of at least --3000~km~s$^{-1}$ from rest \citep{wmfh91}.  If no trough meets these criteria, the $BI \equiv 0$.  BAL QSOs are defined as those QSOs which have $BI > 0$ for some absorption line.

The constraint of a detachment velocity \hbox{$<$--3000~km~s$^{-1}$} may cause one to overlook some objects with broad absorption in our sample.  We investigate this possibility by defining $BI_0$, which is calculated similarly as $BI$ but integrated to a detachment velocity of 0~km~s$^{-1}$.  We find that we are not missing BAL QSOs, as $BI > 0$ whenever $BI_0 > 0$ for all the sources.  Some BALs do extend beyond a velocity of --3000~km~s$^{-1}$, and $BI_0 > BI$ in these cases.  Further modifications to the traditional $BI$ index were considered by \citet{thrrsvkafbkn06}, but $BI$ and $BI_0$ are sufficient for our purpose of simply identifying BALs.

Our sample of 128 QSOs includes 9 sources which were identified as HF03 BAL QSOs.  Our continuum fits give $BI > 0$ for 8 of these sources (0018+0047, 0051--0019, 0109--0128, 1208+1535, 1235+1453, 1243+0121, 1314+0116, and 1331--0108).  We also include 1234+0122, which was identified as an HF03 HiBAL with a low $BI = 4$.  Our continuum fit technically gives $BI = 0$ for 1234+0122, but multiple strong absorption features are clearly present, with one just under 2000~km~s$^{-1}$ wide.  (The absorption appears to have weakened by the SDSS epoch.)  For three additional sources (0010--0012, 0055+0025, and 1213+0922), we measure $BI > 0$, although the absorption is weak in these sources and they were not classified as BAL QSOs by HF03.  The final source, 0021--0213, was observed in the LBQS and also with the {\it HET}, but not in the SDSS.  Including 1234+0122, we have 13 QSOs in our sample with \ion{C}{4} BALs, i.e., $BI > 0$.  The ratio spectra, $R(v)$, for the BAL QSOs in our sample are shown in Figure~\ref{cIVBALsFig}.  In this figure, the SDSS spectra have been convolved with a Gaussian of the appropriate width to match the resolution of the LBQS spectra.

We have assumed that the continuum in the \hbox{1400--1550}~\AA\ region is a smooth, reddened power law.  In fact, that may not be the case.  While the narrow absorption features in our sample spectra are reasonably attributed to absorption, we cannot exclude the possibility that broad, shallow features are due to structure in the continuum emission.  The most plausible candidate in our sample for this effect is 1213+0922, which has a broad (15,000~km~s$^{-1}$ wide), shallow, wedge-shaped feature with a minimum at about --16,000~km~s$^{-1}$.  This feature is seen in both the LBQS and SDSS epochs.  We cannot rule out the possibility that the feature is due to continuum structure, but if it is, the results of this study would not be significantly impacted.

None of the BAL QSOs in our sample was considered by L07.  One source in our sample, 1331--0108, was observed twice in 1991 by B93.  We do not directly compare our measurements to those of B93 for this object because we cannot account for differences in analysis such as continuum placement.  We include 1331--0108 in both samples when comparing results between studies as if it were an independent source in each study.

The dates of the LBQS and SDSS (or {\it HET}) observations are given in Table~\ref{obsCatTab}.  The measured properties of our sources are listed in Table~\ref{bALDataTab} and Table~\ref{bITab}.  We discuss our BAL QSO sources in detail in the Appendix.

\section{BAL VARIABILITY METRICS\label{bALVarMetricSec}}
In this section, we apply various metrics to characterize and constrain BAL variation quantitatively.

\subsection{Variation in Velocity Space\label{varVelProfSec}}

Figure~\ref{cIVBALsFig} shows the \ion{C}{4} region (1400--1550~\AA) ratio spectra $R(v)$ for the 13 QSOs in our sample.  Clearly, BAL absorption variation occurs frequently on multi-year time scales.  In some cases, the entire BAL varies (e.g., 0109--0128), while in others, only part of the BAL changes (e.g., 0051--0019).  In this section, we define a criterion to identify velocity regions for which BAL absorption has varied between epochs.  We consider the typical width of a varying region (\S\ref{charVardVSec}), the strength of variation in these regions (\S\ref{strVarSec}), and the dependence of variation on the overall absorption strength in a region (\S\ref{depVarStrVarSec}).  In order to compare the LBQS, SDSS, and {\it HET} spectra directly, we rebinned all spectra to a common grid with bins 1~\AA\ wide.  This binning oversamples the LBQS resolution by $\approx$2--3 and conveniently allows comparisons between sources and across epochs on a common velocity-space grid.  For the studies in this section only, we smoothed the observed-frame LBQS spectrum with a boxcar three bins wide before rebinning in order to reduce scatter without significantly degrading resolution, and convolved the SDSS spectrum with a Gaussian in order to match the resolution of the LBQS ratio spectra, which is $\approx$3~\AA\ in the rest frame.  In this section, we use metrics which are relatively insensitive to the measurement errors in individual bins.

\subsubsection{Characteristic Velocity Width of Variation\label{charVardVSec}}

To identify a set of velocity regions in each BAL where variation has occurred, we determine the regions in the \ion{C}{4} absorption spectrum of each BAL where the ratio spectra, $R(v)$, differed by at least 0.1 over 1200~km~s$^{-1}$ (6 bins) or more between epochs.  In bins where the ratio spectrum rose above the continuum ($R(v) > 1$), we pegged the value of the ratio spectrum to 1 in order to minimize the effects of statistical noise and emission structure we have not modeled.  While our definition of ``varying region'' allows for cases where $\Delta R(v) > 0.1$ in some bins and $\Delta R(v) < 0.1$ in other bins of the same region, there are only two instances where this actually occurs (in 1235+1453 and 1314+0116).  This indicates that the identified regions are not strongly influenced by noise in the spectrum.

\begin{figure} [ht]
  \begin{center}
      \includegraphics[width=2.3in, angle=270]{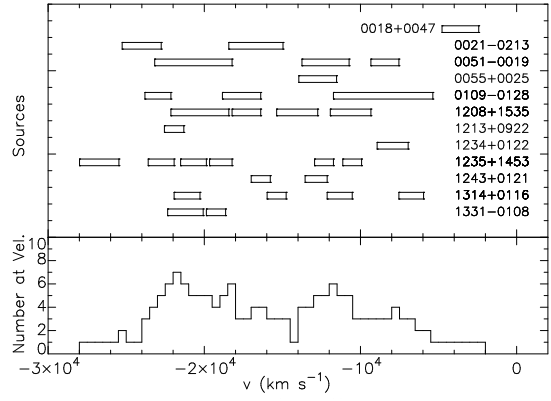}
      \caption{\label{vRegFig}{\it Top panel:}  The horizontal bars indicate velocity regions which were observed to vary (in the sense of \S\ref{charVardVSec}) for the sources listed.  {\it Bottom panel:}  The number of times a particular velocity was included in a varying region.}
   \end{center}
\end{figure}

We identify 30 varying regions, and 12 of our 13 sources have at least one varying region.  The regions are listed in Table~\ref{varRegTab} and are indicated on the ratio spectra in Figure~\ref{cIVBALsFig}.  In Figure~\ref{vRegFig}, we display the distribution of varying regions together with a histogram of the number of times a particular velocity was included in a varying region.  As the figure shows, variation occurs in our sample with roughly equal probability from --6000 to --24,000~km~s$^{-1}$.

This definition of ``varying region'' does not identify all types of BAL variation.  For example, it does not flag the trough contraction in 0010--0012, and it is not sensitive to acceleration in narrow troughs.  It is intended to be a well-defined, reliable way to flag significant variation without referring to the shapes of individual absorption troughs.  Averaged across all sources, LBQS error spectrum is about 0.1 in a 1~\AA\ bin.  Given that the smallest varying regions we identify are 6 bins wide, it is unlikely that many of them are due to statistical fluctuations.

\begin{figure} [ht]
  \begin{center}
      \includegraphics[width=2.3in, angle=270]{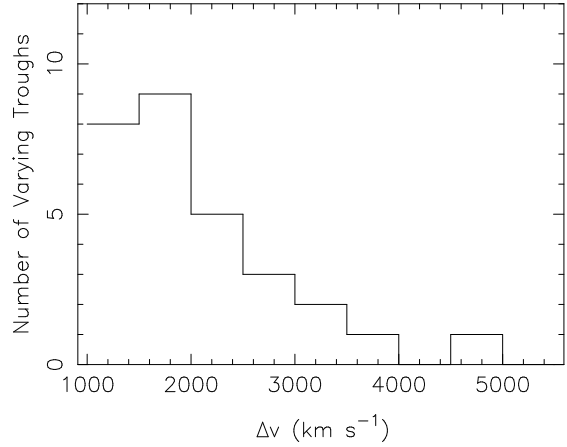}
      \caption{\label{dVHistFig}The number of regions in all BAL QSO sources with width $\Delta v$ that varied between epochs.  We only consider varying regions with widths $\ge$1200~km~s$^{-1}$ wide in order to minimize false detections due to statistical variation.  We do not consider varying regions of smaller widths.}
   \end{center}
\end{figure}

Figure~\ref{dVHistFig} shows the number of times variation was observed in a region with a given velocity width.  Variation tends to occur on small ($\la$2000~km~s$^{-1}$) velocity scales.  Even the largest variation widths (5000~km~s$^{-1}$) are narrow compared to the extent of strong BAL absorption.  The velocity width of a varying region is not significantly correlated with the outflow velocity of that region, according to a Spearman rank correlation test.

In a few cases, two varying regions are sufficiently close together to raise the possibility that they are separated only because statistical noise obscured the variation in the intervening bins.  Even if these regions were merged, it would not affect our conclusion that variation occurs on relatively small velocity scales.

We define $\Sigma_{\Delta v}$ as the sum of velocity widths of all varying regions in a single source.  For example, if one BAL QSO had two varying regions of velocity width 2000 and 3000~km~s$^{-1}$, then $\Sigma_{\Delta v} = 5000$~km~s$^{-1}$ for that QSO.  The values of $\Sigma_{\Delta v}$ for our objects are listed in Table~\ref{varRegTab}.  We do not find a significant correlation between $\Sigma_{\Delta v}$ and $\Delta t_{sys}$, the rest-frame time between epochs, but the range of $\Delta t_{sys}$ covers only a factor of $\approx2$ in our sample.

\subsubsection{Strength of Variation\label{strVarSec}}

\begin{figure} [ht]
  \begin{center}
      \includegraphics[width=2.3in, angle=270]{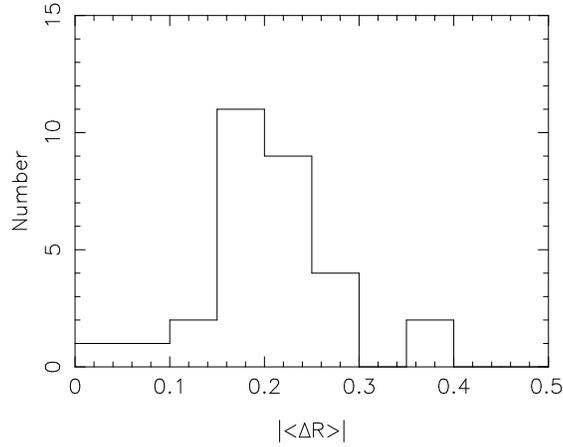}
      \caption{\label{dRHistFig}The distribution of $|\langle\Delta R\rangle|$, the average change in the ratio spectrum between epochs in regions at least 1200~km~s$^{-1}$ wide which varied (as defined in \S\ref{charVardVSec}).  Our definition of ``varying region'' is not sensitive to very shallow absorption variation, so there may be additional variation with small $|\langle \Delta R \rangle|$ that we have not identified.}
   \end{center}
\end{figure}

For each velocity region determined to vary (\S\ref{charVardVSec}), we determine the magnitude of the mean difference, $|\langle\Delta R\rangle|$, between the ratio spectra in that region.  (Here $\Delta R(v)$, the change between epochs in the bin at velocity $v$, is averaged over all velocity bins in a varying region.)  Figure~\ref{dRHistFig} shows the distribution of $|\langle\Delta R\rangle|$.  Because we imposed a threshold of $|\Delta R| \ge 0.1$ for a bin to be considered part of a varying region, we do not expect many cases with $|\langle\Delta R\rangle| < 0.1$.  (By our definition, a varying region of contiguous bins could, in principle, have $|\langle \Delta R\rangle| < 0.1$ if $\Delta R > 0.1$ in some bins and $\Delta R < -0.1$ in others; this happens in only two cases.)

The distribution of $|\langle\Delta R\rangle|$ is strongest at 0.15--0.25, indicating that changes in absorber depth of $|\langle\Delta R\rangle| \la 0.25$ are favored on multi-year time scales.  We find no significant correlation between $|\langle\Delta R\rangle|$ and the velocity width of the varying region or the average velocity of the varying region.  $\langle\Delta R\rangle$ is positive for 13 of 30 varying regions.  Surveys with larger samples will be able to constrain further any positive (negative) bias in the distribution of $\langle\Delta R\rangle$ which could arise if BALs strengthen and weaken on different time scales.
 
\subsubsection{Dependence of Variation on Absorption Strength\label{depVarStrVarSec}}
We would like to determine whether the properties of variation are dependent on the depth of the absorption in the varying region.  To do this, we consider three distributions.  The first, $R_{all}$, is defined as the distribution of the ratio spectrum $R(v)$ (averaged between epochs) in every velocity bin from --30,000 to 0~km~s$^{-1}$ for all the sources in our sample.  $R_{all}$ roughly represents the distribution of absorption depths in all velocity bins of the ``average BAL'' in our sample.

We then compare $R_{all}$ to the distribution of the ratio spectra for the more-absorbed epoch and also for the less-absorbed epoch.  We define $\langle R_< \rangle$, as the ratio spectrum, $R(v)$, averaged over the bins in a varying region in the epoch for which absorption is stronger (and therefore $R(v)$ is smaller).  We take the more-absorbed epoch for each individual region, and for some QSOs with multiple regions (e.g., 1208+1535), this means we take the average ratio spectrum in the first epoch for some regions and in the second epoch for other regions.  Similarly, we define $\langle R_> \rangle$, obtained from the epoch with weaker absorption (and therefore a larger average $R(v)$) for each region.

\begin{figure} [ht]
  \begin{center}
      \includegraphics[width=2.3in, angle=270]{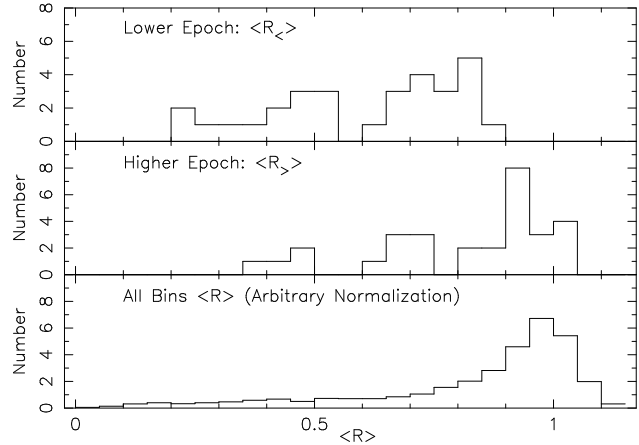}
      \caption{\label{avgRHistFig}{\it Top panel:}  The distribution of $\langle R(v) \rangle$ in the more deeply-absorbed epoch.  {\it Middle panel:}  The distribution of $\langle R(v) \rangle$ in the less-absorbed epoch.  {\it Bottom panel:}  The distribution of $\langle R(v) \rangle$ in all 1~\AA\ bins from 1400--1550~\AA\ for both epochs of all sources in our sample.  The normalization of the bottom panel is arbitrary.  See \S\ref{depVarStrVarSec} for more information.}
   \end{center}
\end{figure}

In Figure~\ref{avgRHistFig}, we plot the distributions of $\langle R_< \rangle$ (top panel), $\langle R_> \rangle$ (middle panel), and $R_{all}$ (bottom panel, with arbitrary normalization).  The distribution of absorption depths in the varying regions in the epoch of weaker absorption ($\langle R_> \rangle$) resembles that of the average BAL absorption ($R_{all}$), while the distribution from the more strongly-absorbed epoch ($\langle R_< \rangle$) differs considerably from $R_{all}$.  This may simply be a consequence of the fact that most bins in the 1400--1550~\AA\ region are not strongly absorbed for the average BAL, and the distribution $R_>$, taken from the less-absorbed epoch, more accurately reflects this.

The average velocity of varying regions with $\langle R_> \rangle > 0.9$ is --19,000~km~s$^{-1}$, compared to --13,000~km~s$^{-1}$ for those with $\langle R_> \rangle \le 0.9$.  A Kolmogorov-Smirnov (KS) test suggests (at 99.3\% confidence) that the velocities for these two sets of varying regions are distributed differently.  It is not surprising that varying regions which are weakly absorbed in one epoch tend to appear at higher velocities, as BAL absorption is, on average, weaker at higher velocities \citep[e.g.,][]{kvmw93}.

A total of 20 of the 30 varying regions have $\langle R_< \rangle \ge 0.5$, indicating a preference for measurable variation to occur in regions with weaker absorption.  Half of the varying regions have $\langle R_> \rangle \ge 0.9$, corresponding to weak or even no absorption in one epoch.  In the $\langle R_{all} \rangle$ distribution, 88\% of bins have $\langle R \rangle \ge 0.5$, and 57\% have $\langle R \rangle \ge 0.9$.  The tendency for variation to occur in weakly-absorbed regions is therefore a reflection of the fact that most bins in our sample spectra are weakly absorbed, combined with the tendency for variation to occur at a wide range of outflow velocities.  We find no evidence for a correlation between the strength of variation in a varying region, $|\langle\Delta R\rangle|$, and the average depth of that region, $\langle R\rangle$.  We therefore find no evidence that the probability of variation in a velocity region is dependent on the absorption strength in that region.

\subsubsection{Fraction of Absorption Width That Varies\label{fracAbsVarSec}}

We define $f_{BAL}$ as the fraction of the 1~\AA\ wide wavelength bins from 1400--1550~\AA\ which are at least 20\% absorbed ($R(\lambda) \le 0.8$) in one or both epochs.  We determine the fraction of varying wavelength bins, $f_{vb}$, by counting the number of 1~\AA\ bins from 1400--1500~\AA\ which vary by at least $|\Delta R(\lambda)|=$0.2 between epochs.  (Here we require a greater threshold of change $|\Delta R(\lambda)|$ than in \S\ref{charVardVSec} because we are considering single bins.)  Because all spectra are binned onto the same 1~\AA\ grid which has 149 bins, the {\it fractions} $f_{BAL}$ and $f_{vb}$ are both proportional to the {\it number} of deep or varying 1~\AA\ bins, with the same constant of proportionality, 149.  We do not cap the ratio spectra at 1 when determining $f_{vb}$, as this could bias our results.  We have not accounted for absorption from intervening systems, so QSOs with strong intervening system lines in the \ion{C}{4} region may have a slightly higher $f_{BAL}$ than measured from the BAL alone.  

\begin{figure} [ht]
  \begin{center}
      \includegraphics[width=2.3in, angle=270]{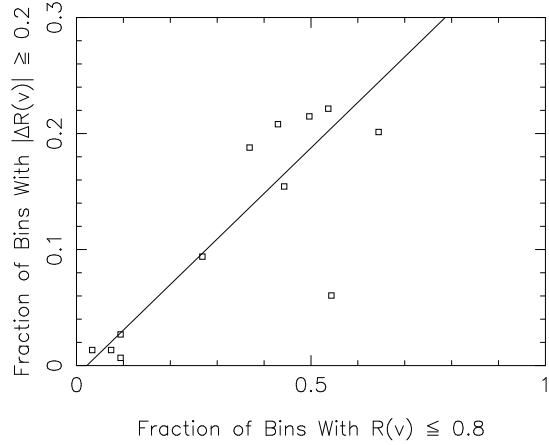}
      \caption{\label{fBBFVBFig}The fraction of 1~\AA\ bins in the \ion{C}{4} absorption region with ratio spectra varying by $|\Delta R(\lambda)| \ge 0.2$ is correlated with the fraction of BAL bins ($R(\lambda) \le 0.8$) in that region.  The best linear fit (Equation~\ref{fBBFVBEqn}) is shown.}
   \end{center}
\end{figure}

A Spearman rank correlation test finds that $f_{vb}$ and $f_{BAL}$ are correlated at 99.9\% confidence.  That is, BALs with broader regions of deep absorption ($R(\lambda) \le 0.8$) tend to have a larger number of variable wavelength bins from rest to 30,000~km~s$^{-1}$.  A linear fit gives:
\begin{eqnarray}
f_{vb} &=& -0.01 \pm 0.03 + (0.39 \pm 0.07) f_{BAL}.\label{fBBFVBEqn}
\end{eqnarray}
The errors are estimated from the scatter in the data (rather than from unknown error on individual data points), assuming $\chi^2_{\nu} = 1$; see pp. 666--669 of \citet{ptvf02}.  Figure~\ref{fBBFVBFig} shows the plot of $f_{vb}$ against $f_{BAL}$ for our sources together with the fit from Equation~\ref{fBBFVBEqn}.

$f_{BAL}$ and $\Sigma_{\Delta v}$ (defined in \S\ref{charVardVSec}) are correlated at the 98\% confidence level, supporting the finding that broader (deep) BALs have more varying absorption bins.  If we confine the search for varying bins to only include those which are deeply absorbed ($R(\lambda) \le 0.8$), we find a similarly strong correlation (at 99.8\% confidence) as varying bins primarily occur in the strongly-absorbed regions.

\subsection{EW Variation}

In \S\ref{varVelProfSec}, we determined ranges of velocity bins over which the BAL was observed to vary beyond a certain threshhold.  The strength of the variation in a particular velocity bin was not important, so long as it exceeded the threshold.  By contrast, tests for variation in EW also account for changes in absorption depth.  We calculate EW errors formally, estimating a continuum error of 5\%.

\subsubsection{Cumulative EW Variation\label{cumEWVarSec}}

\begin{figure} [ht]
  \begin{center}
      \includegraphics[width=2.3in, angle=270]{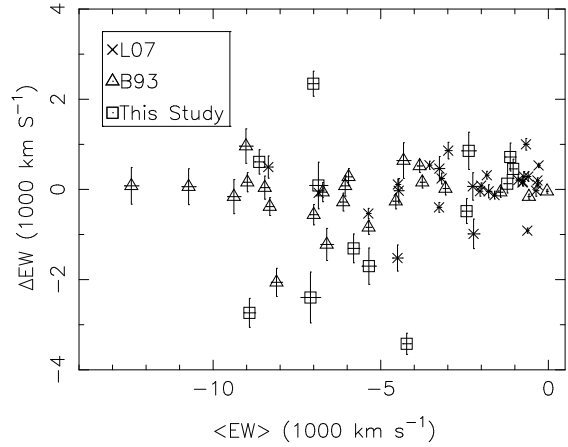}
      \caption{\label{totdEWVsAvgtotEWsFig}$\Delta EW$ vs. $\langle EW \rangle$ (in~km~s$^{-1}$) across the entire \ion{C}{4} absorption region for each BAL in our sample.  The symbols described in the legend indicate the samples of B93, L07, and this study.}
   \end{center}
\end{figure}

In Figure~\ref{totdEWVsAvgtotEWsFig}, we plot the total change in equivalent width between epochs, $\Delta EW$, against the average EW, $\langle EW \rangle$, for the entire \ion{C}{4} region of each BAL QSO in our sample.  For comparison, we have also plotted the results from the studies of B93 and L07, which apply to shorter time scales.  In the four cases where L07 measured EWs from two BALs in a single QSO separately, we have combined the EWs to get an approximate EW for the entire \ion{C}{4} absorption region.  In cases where B93 observed a source more than twice, we used data from the two epochs with maximal time separation.

We do not find significant correlations between $\langle EW \rangle$ and $\Delta EW$ in the combined sample.  There is a significant ($>$99.99\% confidence) correlation in the combined sample between $\langle EW \rangle$ and the magnitude of fractional change in EW, $|\Delta EW / \langle EW \rangle|$, so that the magnitude of fractional variability is greater for weaker BALs.  However, this correlation is primarily driven by the weakest BALs.  It disappears (confidence $<$ 90\%) if we exclude the 20\% of BALs (mostly taken from L07) in the combined sample which have EWs $>$--1000~km~s$^{-1}$.  For this reason, we do not draw strong conclusions from the correlation.

\begin{figure} [ht]
  \begin{center}
      \includegraphics[width=2.3in, angle=270]{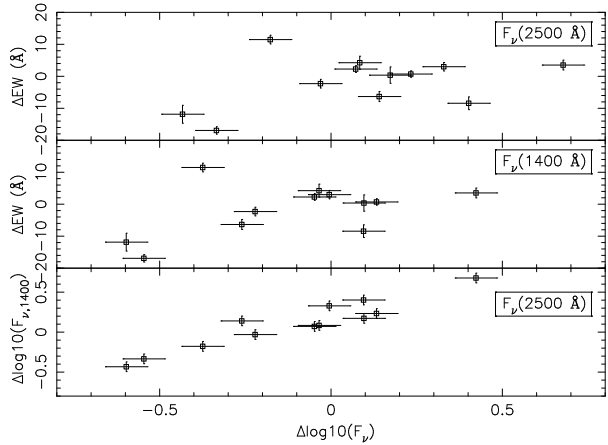}
      \caption{\label{dCIVEWdLFig}The change in \ion{C}{4} region EW as a function of the change in flux.  The error on the flux is assumed to be dominated by the LBQS flux error, estimated to be about 15\% \citep{hfcfwmam91}.  The EW errors are calculated from the formal error on EW, assuming the continuum error is 5\%.  {\it Top panel}:  the change in flux, $\Delta\log(F_{\nu})$, is calculated at 2500~\AA.  {\it Middle panel}:  $\Delta\log(F_{\nu})$ is calculated at 1400~\AA.  {\it Bottom panel}:  $\Delta\log(F_{\nu})$ at 1400~\AA\ ($y$-axis) is plotted against $\Delta\log(F_{\nu})$ at 2500~\AA\ ($x$-axis).  The two are strongly correlated.}
   \end{center}
\end{figure}

We plot $\Delta EW$ against the logarithm of the change in flux density at 1400 and at 2500~\AA\ in Figure~\ref{dCIVEWdLFig}.  We find no significant correlation between the change in flux density, $\Delta F_{\nu}$, and $\Delta EW$, or between $|\Delta F_{\nu}|$ and $|\Delta EW|$, as might be expected in a photoionization-induced variation scenario.  Ionization fractions for a given ion are roughly symmetric about a peak at some ionizing flux (for a given shape of the ionizing continuum).  While the ionization fraction for a given ion can remain unchanged if the ionizing flux changes by just the right amount, this is a contrived scenario, and is unlikely to occur in a large number of cases.  We discuss physical causes of EW variation further in \S\ref{photoVarSec} and \S\ref{covFactVarSec}.

The bottom panel of Figure~\ref{dCIVEWdLFig} shows that the changes in $F_{\nu}$ at 1400 and 2500~\AA\ are correlated.  The photoionization of \ion{C}{4} is driven by photons at higher energies ($\ga$50~eV) than we have observed.  If the high-energy continuum drives photoionization variation, it would have to vary independently of the observed continuum (at $\lambda \ge 1400$~\AA).  The nature of QSO continuum variation from the UV to X-rays is not currently well-constrained; however, this phenomenon has received some attention for lower-luminosity Seyfert galaxies.  In the case of the Seyfert 1 galaxy NGC~3516, the X-ray and optical continua have been observed to vary independently \citep{mmen02}.  Studies of other Seyfert 1 galaxies have found time lags of tens of minutes to days between the X-ray and optical continua \citep[e.g.,][]{sunm03, dfklmglmsbhnop06}.

\subsubsection{EW Variation at Different Outflow Velocities\label{eWVarDiffVsSec}}

In order to test for any velocity-dependence in BAL variation, we calculate the EW for each BAL in 5000~km~s$^{-1}$ wide bins from --30,000 to 0~km~s$^{-1}$.  We find no evidence of correlation between $\Delta t_{sys}$ and either $\Delta EW$ or $|\Delta EW / \langle EW \rangle|$ in any of the velocity bins.  However, the range of $\Delta t_{sys}$ in our sample covers only a factor of $\approx$2, so we are not sensitive to velocity-dependent variation on a wide range of time scales, nor are we sensitive to faster variation which does not generate trends on multi-year time scales.

BAL absorption is likely saturated with a depth primarily determined by the covering factor \citep[e.g.,][]{ablgwbd99}.  BAL absorption is, on average, shallower at high velocities.  If shallower absorption regions are less saturated, high-velocity absorption could be more responsive to changes in the BAL absorber.  However, we find no indication (either visually or formally with a KS test) that the distribution of $\Delta EW$ changes across velocity ranges.

Similarly, if absorption at higher velocities were less saturated, high-velocity variation may track the continuum more closely.  We test for correlations between $\Delta F_{\nu}$ at 2500~\AA\ and $\Delta EW$ in 5000~km~s$^{-1}$ velocity bins.  The strongest correlation is between $|\Delta EW|$ and $\Delta F_{\nu}$ in the --10,000 to --15,000 km~s$^{-1}$ velocity range.  At 99\% confidence, the putative correlation is not highly significant, given that we have tested 12 cases for correlations (both $\Delta EW$ and $|\Delta EW|$ against $F_{\nu}$ in 6 velocity bins).  We therefore do not find any strong evidence for velocity-dependent correlations with the continuum.

\subsubsection{Evolution of $|\Delta EW|$\label{evoldEWSec}}

\begin{figure} [ht]
  \begin{center}
      \includegraphics[width=2.3in, angle=270]{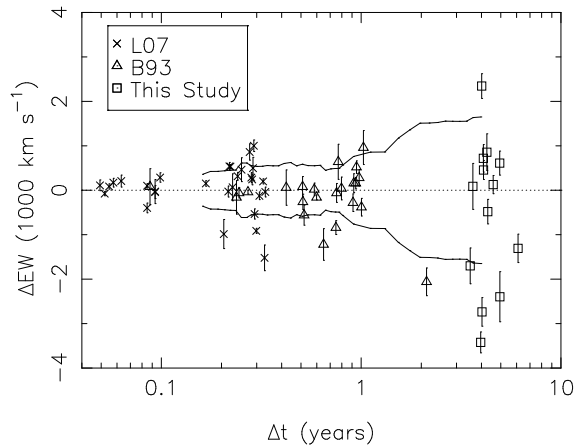}
      \caption{\label{totdEWVsTFig}The change in EW against the rest-frame time between epochs.  The envelope of $\Delta EW$ clearly expands with time.  We show data from B93, L07, and our study, identified with symbols as shown in the legend.  The $x$-axis is logarithmic.  The solid lines indicate the square root of the unbiased sample variance calculated from a sliding window of 15 time-ordered entries.}
   \end{center}
\end{figure}

In order to compare the L07 sample to the BALs in our study, we construct a matched sample of L07 objects which have EWs greater in magnitude than that of the weakest BAL in our sample.  We call this set of 18 objects the ``strong L07 BALs.''  Figure~\ref{totdEWVsTFig} shows the change in EW, $\Delta EW$, plotted against the time between epochs.  The data from B93 and the strong L07 BALs are included.  The range (or ``envelope'') of $\Delta EW$ is clearly increasing with time.  The B93 BALs were slightly stronger on average than our BALs, but had smaller $|\Delta EW|$ over shorter time scales.  For the strong L07 BALs, the mean $|\Delta EW|$ of 390~km~s$^{-1}$ is much smaller than that of our sample (1300~km~s$^{-1}$).

In Figure~\ref{totdEWVsTFig}, we plot the standard deviation of a sliding window of 15 time-ordered objects to illustrate the increasing spread of $\Delta EW$.  The mean value of $|\Delta EW|$ in our sample is 1300~km~s$^{-1}$ over a mean time of 4.3~yr, while the standard deviation is $\approx$1650~km~s$^{-1}$ at $\Delta t_{sys} \approx 4$~yr.

A strong BAL with $\langle EW \rangle = -6000$~km~s$^{-1}$ which varies at the rate of 400~km~s$^{-1}$~yr$^{-1}$ would have a characteristic lifetime of $|\langle EW \rangle / (\Delta EW / \Delta t_{sys})| \approx15$~yr.  If the envelope surrounding $\Delta EW$ keeps growing on longer time scales, some strong BALs could completely disappear over a few decades.

We might also expect that strong BALs would appear in QSOs that had previously shown no BALs.  (We discuss constraints on BAL transience further in \S\ref{bALTransienceSec}.)  Future studies with a large number of BAL observations over a wide range of time scales will allow careful study of the time-evolving distribution of $\Delta EW$, which is apparently revealing important physics of the QSO environment.

\subsubsection{Evolution of $|\Delta EW / \langle EW \rangle|$\label{evolFracdEWTimeSec}}

BALs have been observed to vary on short (multi-month) time scales (\S\ref{indivBALVarSec}, \S\ref{bALVarSampleSec}), and L07 found that the largest fractional variability $|\Delta EW / \langle EW \rangle|$ was seen in sources with the longest ($\approx4$ month) rest-frame time between epochs.  If BAL QSOs evolve monotonically over several years, we would expect to see larger variation on the longer time scales in our study.

The mean value of $|\Delta EW / \langle EW \rangle|$ is $0.33 \pm 0.04$ for all L07 objects and $0.32 \pm 0.04$ for our sample.  However, the mean of the L07 sample is dominated by the three outlier sources with $|\Delta EW / \langle EW \rangle| \approx1.5$, and all three outliers are from sources with weaker BALs than in our sample.  For the strong L07 BALs alone (\S\ref{evoldEWSec}), the mean is $0.12 \pm 0.02$.  For the B93 sources, the mean fractional change in EW is only about $0.12 \pm 0.06$, similar to that of the L07 sources with stronger BALs.  A KS test indicates that $|\Delta EW / \langle EW \rangle|$ is distributed differently between the B93 sources and our sources at $>$99.9\% confidence.  The distribution of fractional change in EW also varies between the strong L07 BALs and our sample at 99.3\% confidence.

\begin{figure} [ht]
  \begin{center}
      \includegraphics[width=2.3in, angle=270]{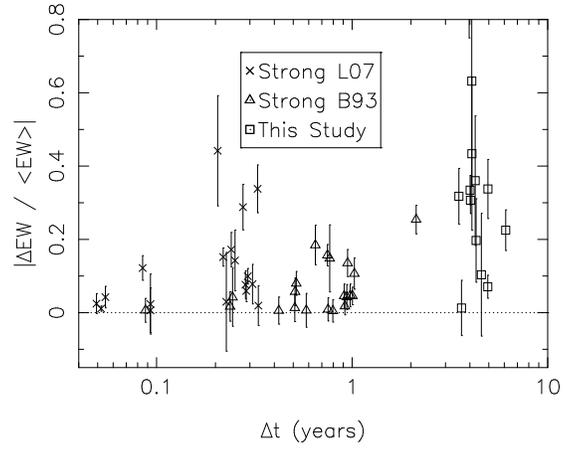}
      \caption{\label{fracdEWVsTFig}The magnitude of fractional change in EW against the rest-frame time between epochs for sources in our study, as well as for those of B93 and L07.  We have constructed a matched sample of sources by including only the B93 and L07 sources with $|\langle EW \rangle|$ greater than the minimum in our sample.  The $x$-axis is logarithmic.  The symbols described in the legend indicate the samples of B93, L07, and our sample.}
   \end{center}
\end{figure}

We have plotted our values of $|\Delta EW / \langle EW \rangle|$ against time in Figure~\ref{fracdEWVsTFig} along with a matched sample of values from B93 and L07.  We have only included sources with $|\langle EW \rangle|$ greater than the minimum of our sample in the plot.

In summary, we find that (apart from a few outliers) the variation in fractional EW is greatest over the longest times.  For three outliers with relatively small $\langle EW \rangle$ in the L07 sample, $|\Delta EW / \langle EW \rangle|$ has varied greatly over short (2--4 month) times.

\subsection{Acceleration at Detachment Velocity}

The relatively long time scales in our study allow a sensitive search for acceleration of BAL components.  Previous studies have found accelerations of $\approx$0.03~cm~s$^{-2}$ for Q~1303+308 \citep{vi01} and $\approx$0.08~cm~s$^{-2}$ for Mrk~231 \citep{rvs02}.  \citet{hsher07} have recently presented evidence for acceleration in a \ion{C}{4} BAL trough of SDSS~J024221.87+004912.6 at a rate of $\approx$0.15 cm s$^{-2}$ over 1.4~yr.  The detachment velocity of this BAL trough increased in magnitude between epochs by $\approx$70~km~s$^{-1}$.  The wavelength calibration of the LBQS spectra is accurate to about 4~\AA\ \citep{fchmtwa87}, corresponding to about 260~km~s$^{-1}$ at 1549~\AA\ for a $z = 2$ QSO.  If the BALs in our sample accelerate over 6~yr at the same rate as seen in SDSS~J024221.87+004912.6, we should detect noticable trough changes.

The variation in SDSS~J$024221.87+004912.6$ was most evident in the onset region of the deep \ion{C}{4} BAL.  The objects in our sample with deep BAL troughs generally have steep onset regions (at $\approx$--5000~km~s$^{-1}$) which are ideal for tests of acceleration (Figure~\ref{cIVBALsFig}).  After convolving the SDSS spectrum with a Gaussian to approximate the LBQS resolution, we do not find strong candidates for acceleration in the BAL onset region.  This does not necessarily mean that material is not accelerating along the outflow; the onset region may be continually replenished with material at a constant velocity.  We discuss the physical implications of this measurement in \S\ref{accelSec}.

The ratio spectrum, $R(\lambda)$, drops below a threshold of $R(\lambda) =$0.3 at the same wavelength (within 1~\AA) in both epochs for the 7 sources listed in Table~\ref{accelLimitTab}.  The upper limits on acceleration ($\Delta v / \Delta t_{sys}$) for the seven sources with sharp BAL onsets range from 0.12 to 0.17~cm~s$^{-2}$ and are given in the table.  The acceleration limits were calculated using the different rest-frame time elapsed between epochs for each source, assuming an upper limit of 1~\AA\ for movement of the absorption region.  In 3 of the 7 sources (0018+0047, 0051--0019, and 1331--0108), narrow absorption lines from apparent intervening systems are well-matched between epochs and further support the use of 1~\AA\ as a constraint on the relative wavelength calibration.\footnote{Shifting the narrow absorption regions of the SDSS ratio spectrum in wavelength space and fitting against the LBQS ratio spectrum indicate that the wavelength calibration for these sources may be good to 0.5~\AA\ or less.  However, unmodeled emission and systematics prevent us from using the $\chi^2$ statistic to place strong confidence limits on the relative wavelength calibration.}  The velocities of the intervening systems are given in Table~\ref{bALDataTab}.

An eighth source, 1208+1535, also has a steep onset region, but the BAL variation is so complex that it is difficult to determine whether acceleration is a factor.  Shifting the later (SDSS) epoch spectrum to lower velocities by about 2500~km~s$^{-1}$ would match some features in the LBQS-era BAL trough, but other features would not match well.  This QSO is an excellent target for future monitoring.

The variation in most absorbed regions of the sources in our sample (excepting the onset region of strong BALs) is too complex to test for acceleration.  Visual inspection of weaker absorption regions shows some variation between epochs, including narrowing of the absorption region (e.g., in 0010--0012 at $\approx$--13,000~km~s$^{-1}$) or variation on only one side of the trough (e.g., 1234+0122 at $\approx$--13,000~km~s$^{-1}$).  Only the lowest-velocity component of 0109--0128 (at $\approx$--3000~km~s$^{-1}$) shows visual evidence of acceleration on both sides of the absorption trough.  Unfortunately, in this case there are no narrow lines from intervening or other intrinsic systems strongly present in both epochs which can be used to test the relative wavelength calibration.  The doublet structure of the feature (seen in the full resolution of the SDSS spectra) indicates that the feature, if intrinsic, could be attributed to \ion{C}{4} flowing outward at --3400~km~s$^{-1}$.  We do not see any clear cases of significant absorption from lines of other ions at this outflow velocity in either the LBQS or SDSS spectrum.  Future spectroscopy of this object will be able to determine whether or not this is an accelerating, intrinsic absorption component.  If this feature is actually accelerating, and continues to accelerate at the same rate, it will join the main BAL trough in about 70 yr (in the observed frame).

\section{DISCUSSION\label{sumAndDiscSec}}

In the following sections, we consider the physical implications of our study.  First, we present a brief summary of the main quantitative results of our analysis in \S\ref{varMetricSumSec}.  We then briefly discuss constraints that our study places on BAL transience (\S\ref{bALTransienceSec}).  We also consider the implications our study has for the physical processes of photoionization-induced variation (\S\ref{photoVarSec}) and  geometric variation (\S\ref{covFactVarSec}).  Finally, we discuss the implications of our acceleration constraints on the BAL onset region (\S\ref{accelSec}).

\subsection{Summary of Variation Metrics\label{varMetricSumSec}}

Although \ion{C}{4} BAL absorption can extend up to (and even beyond) --30,000~km~s$^{-1}$ and BAL troughs can be tens of thousands of km~s$^{-1}$ wide, variation tends to occur in (multiple) narrower bands of width $<$5000~km~s$^{-1}$, with smaller widths ($\la$2000~km~s$^{-1}$) most common.  The varying regions of all sources combined are distributed across a wide range of velocities.  We find no evidence that the velocity width of a varying region is correlated with the outflow velocity of that region.

The degree of variation, measured in terms of the average change in the ratio spectrum across the varying regions, $|\langle\Delta R\rangle|$, peaks at $|\langle\Delta R\rangle| \approx0.2$.  The algorithm used to determine varying regions generally requires that $|\langle\Delta R\rangle| \ga 0.1$ (set by the data quality), so we are not able to identify weak, broad variation.  The degree of variation $|\langle \Delta R \rangle|$ is not strongly correlated with the absorbed depth, velocity, or velocity width of outflowing regions, nor with the elapsed time between observing epochs.

About half of the varying regions are associated with components which are seen strongly in only one of the two epochs.  These ``transient'' regions tend to appear at higher outflow velocities (--19,000~km~s$^{-1}$) than the non-transient regions (--13,000~km~s$^{-1}$).  This is likely a consequence of the fact that BAL absorption tends to be shallower at higher velocities, while variation occurs across the entire BAL trough.

The number of bins which vary in the \ion{C}{4} absorption region is correlated with the number of deeply-absorbed bins.  The average value of $|\Delta EW / \langle EW \rangle|$ is $\approx0.3$ for our sources, somewhat larger than for a matched sample of L07 and B93 sources observed over shorter time scales.  We do not find evidence that EWs vary differently at different velocities, even though the average BAL absorption varies strongly with velocity \citep[e.g.,][]{kvmw93}.  As Figure~\ref{totdEWVsTFig} clearly shows, the distribution of $\Delta EW$ broadens with time.

We do not find any strong evidence of BAL absorber acceleration in our sample up to wavelength calibration limits, although the variation was so complex in several cases that it could have disguised acceleration.  The steep absorption dropoff at the detachment velocity did not appear to change velocity between epochs.  We found upper limits to acceleration of the onset region of $0.12$ to $0.17$~cm~s$^{-2}$.  The absorbing material {\it in the onset region} is being replenished in such a way as to generally preserve the trough shape, is accelerating intermittently, or is generally accelerating less rapidly than in the case of SDSS~J024221.87+004912.6.  We did identify ambiguous cases for acceleration which we could not constrain, but will test with future observations.

\subsection{BAL Transience\label{bALTransienceSec}}

The degree of variation observed in our study raises the possibility that some strong BALs could be transient on observable time scales (\S\ref{evoldEWSec}).  We briefly present constraints on BAL lifetimes in this section, although the limited nature of our sample, both in numbers and in time scale, prevents us from drawing strong constraints.  Future studies will obtain improved results from large samples of BAL QSOs in surveys like SDSS.

Suppose BAL outflows are a common, but transient, phenomenon in QSOs.  If a given QSO with lifetime $t_{QSO}$ hosts one BAL covering the entire QSO sky with a lifetime $t_{BAL} \ll t_{QSO}$, and if we take two observations of the QSO separated a time $t_{obs} \ll t_{BAL}$ apart, then we have a chance $\approx t_{obs} / t_{QSO}$ of catching the BAL as it appears, assuming the BAL forms quickly.  If a QSO hosts $N_{BAL}$ such BALs over its lifetime (but at most one BAL at a given time), we can detect a BAL appearing during a fraction $\beta \equiv N_{BAL} t_{obs} / t_{QSO}$ of the QSO lifetime.  In a simplified evolutionary scenario in which each QSO hosts BALs for 20\% of its lifetime \citep[based on the frequency of BALs in QSOs;][]{bwgbla00, hf03}, we have $N_{BAL} t_{BAL} = 0.2 t_{QSO}$.  In that case, $\beta = 0.2 t_{obs} / t_{BAL}$.  We observed 115 QSOs with no significant BALs in either epoch.  If BALs are equally probable to appear at any time, then $\beta$ is also the probability that we observe a BAL forming in a given QSO, and a binomial distribution indicates there is a $\ge 90$\% chance we would have seen one BAL appear out of 115 chances if $\beta > 0.02$.  In this scenario, the BAL lifetime is $t_{BAL} > 43$~yr and the QSO lifetime is $t_{QSO} > 210 N_{BAL}$~yr.

If a strong BAL has a lifetime $t_{BAL}$, we would expect BALs observed in the first epoch to disappear in a fraction $t_{obs} / t_{BAL}$ of the observations.  Given 9 QSOs determined to have strong BALs ($BI_0 > 100$) during a first observation epoch, if we make a second observation $t_{obs} \approx$4.3~yr later and the BAL lifetime is $\la$18~yr, we have a $\ge$90\% chance of observing at least one case in which a strong BAL disappears (using a binomial distribution as before).  We therefore expect $t_{BAL} > 18$~yr.  BAL lifetimes may, of course, be at least several orders of magnitude longer.

\subsection{Does Photoionization Variation Drive BAL Variation?\label{photoVarSec}}

Although BAL profiles are likely primarily determined by the velocity-dependent geometric covering factors of saturated absorbers \citep[e.g.,][]{ablgwbd99}, we briefly consider a hypothesis where the trough depth is at least partly determined by the (unsaturated) absorber column density.  In this case, the BAL trough could vary in response to the ionizing continuum.  But, as in previous studies, we have not found any significant correlations between (the magnitude of) absorber variation and continuum variation.  We have tested for correlations with the continuum at both 1400 and 2500~\AA.  Of course, the \ion{C}{4} ionization state will be strongly influenced by the continuum at shorter wavelengths where optical studies do not have spectral coverage.  While the continuum variations at 1400 and 2500~\AA\ are highly correlated, the far UV and X-ray continuua may vary independently of the optical continuum.  As discussed in \S\ref{cumEWVarSec}, this has been observed to happen in at least one Seyfert 1 galaxy.

We find no significant evidence that absorption variation is dependent on absorption depth (\S\ref{depVarStrVarSec}), as we might expect if weakly-absorbed troughs tended to be less saturated (and thus more responsive to changes in ionization levels).  In fact, we observe specific cases where deeply-absorbed regions vary, or contiguous regions at similar absorption depths vary independently (e.g., 0051--0019).

The broadening distribution of $|\Delta EW|$ with time (\S\ref{evoldEWSec}) and the {\it lack} of variation in some parts of BALs is not likely to be due simply to long photoionization time scales.  The gas density (estimated from the \ion{C}{4} recombination time scale) would need to be $<$7000~cm$^{-3}$ to have response times $>$4.2 yr.  For typical equilibrium photoionization models, this would require the BAL absorber to be at least hundreds of parsecs from the ionizing source.  In comparison, the broad emission line regions of QSOs are believed to be less than a few tenths of a parsec from the central source \citep[e.g.,][ and references therein]{ksnmjg00, kbmnss07}.

\subsection{Covering Factor Variation\label{covFactVarSec}}

The observed absorption in BAL troughs is strongly influenced by the absorber covering factor \citep[e.g.,][]{ablgwbd99}.  In this scenario, the absorption may be saturated, but if the absorber does not cover the continuum source completely, the absorption is not black.  In a general sense, covering factor variation also includes scenarios where the continuum is scattered around the absorber \citep[e.g.,][]{ocmtgm99}.

Several factors in our study support a scenario in which the absorber geometry varies.  BALs in our sample vary in discrete velocity segments up to about 5000~km~s$^{-1}$ wide (but typically $\la 2000$~km~s$^{-1}$ wide) which are narrow compared to the extent of a strong BAL.  The change in absorption depth is commonly $|\langle \Delta R \rangle| \approx$0.2 in a varying region.  This leads us to speculate that the BAL absorber contains clumps of material which extend across $\la$2000~km~s$^{-1}$ in radial velocity and cover $\approx$20\% of the continuum emitter (assuming the clumps are optically thick).

Seen individually, these clumps may appear similar to narrower absorption features such as mini-BALs or even NALs.  It has been previously suggested that BAL absorption may be an extension of the NAL phenomenon \citep[e.g., \S4.2.1 of ][]{akdjb99}, and narrower features, including \ion{C}{4} doublets, have been observed to vary \citep[e.g.,][]{hbbbcjl95, gce01, nhbbcjl04, wecg04}.  Modeling is needed to determine if and how it is possible to generate clumps of material which are dynamically linked over (up to) 2000~km~s$^{-1}$ in the BAL outflow.  Previous observations of absorption variation have determined that such structures apparently can exist.  For example, \citet{hbj97} observed intrinsic \ion{N}{5}, \ion{Si}{4}, and \ion{C}{4} absorbers varying in unison over $\la$4 months in the QSO Q$2343+125$.  Each of the \ion{C}{4} doublet components was $\approx$400~km~s$^{-1}$ wide (FWHM), and the overall varying absorption feature was $\approx$1000~km~s$^{-1}$ wide.  The absorbers were found to cover $\la$20\% of the continuum emission, and there was some indication that the variation was driven by changes in covering factor.  Recently, \citet{meck07} observed a mini-BAL with significant substructure that varied in concert in the QSO HS 1603+3820.

Over 3--6 years, about one third of a BAL varies.  Apart from a few outliers, $|\Delta EW / \langle EW \rangle|$ increases with time; it is $\approx$0.3 on average in our sample.  These results are also similar to variation properties of narrower absorption features, as at least 20--25\% of associated absorption lines (AALs; narrow lines within 5000~km~s$^{-1}$ of the emitter rest frame) vary on multi-year time scales \citep[e.g.,][]{nhbbcjl04, wecg04}.

Several specific cases of BAL variation are qualitatively similar to what we would expect from covering factor variation.  We see cases (most notably in 0109--0128) where the change in absorption strength, $\Delta R(\lambda)$, is nearly constant across a wide velocity range.  This could occur in the limiting case where the covering factor is independent of the outflow velocity.  We also see cases (most notably in 0051--0019) where only part of a wide region (all at a common absorption strength $R(\lambda)$) varies.  This could occur if absorbing clumps at some velocities moved out of the line of sight, while clumps of the same size (i.e., covering factor) remained in view at other velocities.  Of course, it is possible to manufacture covering factor scenarios for arbitrary patterns of BAL variation.  Larger samples of BAL variation on a range of time scales would improve our understanding of the general characteristics of BAL variation and would address the plausibility of such scenarios.

Studies with sufficient spectral resolution to resolve doublets in both epochs may also be able to constrain covering factors quantitatively, at least in cases where the absorber has doublet substructures.  The mini-BAL recently monitored by \citet{meck07} contained several narrow sub-components, yet the entire mini-BAL varied in concert.  In this case, the variation was attributable particularly to changes in covering factor.  In another recent study, \citet{hsher07} concluded from an analysis of the doublet structure of \ion{Si}{4} absorption components that the BAL absorbers in SDSS~J024221.87+004912.6 were composed of multiple clouds or filaments with scale sizes $\sim$10$^{15}$~cm.

The fact that the envelope for $\Delta EW$ increases with time (Figure~\ref{totdEWVsTFig}) indicates that BAL outflows are changing along the line of sight on observable time scales.  If the variation is caused by geometric factors such as clumps moving across the line of sight, the pattern of variation is revealing information about the distribution of clumped material in the outflow.  The clumps are apparently common in BALs at velocities ranging from \hbox{--5000} to --25,000~km~s$^{-1}$.  Because we see variation in even the deepest absorption regions, the clumps may not be so numerous at any velocity as to obscure the source several times over (at least in some cases of deep absorption).

\subsection{Acceleration\label{accelSec}}

We searched for acceleration at the onset velocity of strong absorption features previously observed in 7 BALs but found no evidence for such acceleration in our sample.  Over 3--6 yr, acceleration due to radiation pressure on a \ion{C}{4} absorber would certainly be noticable in the sharp absorption onset regions in our sample, unless the absorbers are far enough (parsecs or more) from the emitter to geometrically dilute the radiation.  It has been suggested that the onset region is being replenished (at constant velocity) from the accretion disk, so that the absorption onset velocity does not change even though absorber material is accelerating \citep[e.g.,][]{mcgv95, psk00}.  In this case, the onset region may represent the velocity at which the wind leaving the disk is bent into our line of sight by radiation pressure.

Similarly, the tight upper limit on acceleration of the \ion{C}{4} mini-BAL in the Seyfert 1 galaxy NGC~4151 led \citet{wmgh97} to conclude that the absorber was either very distant from the central source, was slowed by drag in a surrounding medium, or had a complex geometric structure.  A small acceleration upper limit of 0.1~cm~s$^{-2}$ was also found for lines from several ions in the Seyfert~1 galaxy NGC~4051 \citep{kbcer04}.

Over the course of our observations, the accretion disk has rotated significantly.  For a 10$^8$ M$_{\sun}$ black hole, the inner edge of the BAL wind is estimated to be at a radius of $\sim$10$^{16}$~cm \citep{mcgv95}.  Material in a Keplerian orbit at 10$^{16}$ cm (700 gravitational radii) from such a black hole would complete more than two revolutions in 4 yr.  A wind leaving the disk would presumably share the disk rotation.  Given the observed (BAL) structure in the accelerated outflow, the possibility of structure in the accretion disk itself, and variation in the illuminating source, it is remarkable that the onset regions in our sample are so stable.  At a radius of 10$^{17}$ cm, the amount of rotation would be only about 8\% of one revolution.

\section{SUMMARY AND FUTURE WORK\label{concAndFutSec}}

We have searched for \ion{C}{4} BAL variability in QSOs which were observed in both the LBQS and SDSS (or by the {\it HET}).  These surveys were conducted approximately two decades apart (in the observed frame), enabling comparisons between QSOs at $1.7 \le z \le 2.8$ over rest-frame time scales of 3--6 yr.  Because our sources were (with one exception) drawn from two large-scale QSO surveys, we expect that they constitute a reasonably representative sample of optically-selected BAL QSOs in the allowed redshift range.  We summarize our most important findings here.

\begin{enumerate}
\item{BALs tend to vary on multi-year time scales in velocity regions which are a few thousand km~s$^{-1}$ wide, much narrower than a strong BAL.  The varying regions occur at a wide range of outflow velocities and absorption depths.  The typical change in absorption depth is $\la$25\% of the continuum.}
\item{The number of spectral bins which vary over 3--6 yr is correlated with the number of deeply-absorbed bins, and the magnitude of the fractional change in EW, $|\Delta EW / \langle EW \rangle|$, is about 0.3 on these time scales.}
\item{The range of $\Delta EW$ for BAL QSOs increases with time, up to 3--6 yr.  Because we did not see strong BALs appear or disappear, we constrain typical BAL lifetimes to be at least 18~yr.  However, if the envelope of $\Delta EW$ continues to increase on longer time scales, some strong BALs could have lifetimes as short as a few decades.}
\item{We find no evidence that the variation is dominated by photoionization on multi-year time scales.  The variation does not correlate with changes in the observed continuum, although we note that \ion{C}{4} ionization levels are driven by photons at higher energies than we observe.  Even if the ionizing continuum varies independently of the optical/UV continuum, the patterns of variation we observe would require complicated outflow structures.}
\item{Several aspects of the variation we observe are at least qualitatively consistent with a scenario in which the covering factor is changing on multi-year time scales for clumps of material in the BAL outflow.}
\item{We find no evidence for acceleration in the steep onset region of BAL troughs, despite the expected radiation pressure on the absorber.  In scenarios where the onset region is replenished by material flowing off the accretion disk, the disk would have rotated significantly over 4 yr unless the outflow is at a radius $\ga 10^{17}$ cm (for a $10^8$ M$_{\sun}$ black hole).  The flow would have to be remarkably azimuthally smooth not to show evidence of this rotation.}
\end{enumerate}

Previous multi-wavelength studies have searched for correlations with BAL properties.  For example, \citet{gbcpgs06} found a correlation between the maximum BAL outflow velocity and the X-ray weakness for a sample of 35 LBQS BAL QSOs, but did not find evidence for other correlations.  In multi-wavelength studies, the different wavebands are typically observed some time apart.  Our study shows that BALs can evolve significantly over typical time scales between observations, introducing additional scatter into relations between BAL and, e.g., X-ray properties.

Together with the previous work of B93 and L07, we have shown that studies of BAL variability on human time scales yield interesting and important results.  The future of such studies is promising, as the number of known BAL QSOs has increased dramatically with the SDSS.  Based on the results of \citet{thrrsvkafbkn06}, we estimate that the SDSS DR5 QSO catalog contains $\approx$3000 sources at $z \ga 1.7$ with measurable, traditional \ion{C}{4} BALs (having $BI > 0$) and $\approx$7500 with \ion{C}{4} absorption features at least 1000~km~s$^{-1}$ wide.  Observations of these sources with large-scale, spectroscopic surveys will an increase in the size of BAL variability surveys by orders of magnitude, and will enable more sensitive studies of evolution on a range of time scales.

QSOs must be significantly redshifted in order to move the \ion{C}{4} absorption region into the optical spectrum.  At $z = 2$, covering a time range only three times longer than the 3--6 yr span in our study would take nearly a typical professional lifetime (27--54~yr).  It will therefore be difficult to increase the time scale for \ion{C}{4} BAL variation studies greatly unless UV spectra of lower-redshift QSOs are obtained.  There is also much to be gained by improving the coverage on time scales $<$10 yr.  The distribution of $\Delta EW$ and the relation between BALs and narrower absorption features will be interesting to map out for a range of time scales and intrinsic QSO properties.  Increased spectral resolution will also assist in determining absorber substructure and constraining covering factors.

\acknowledgements
We thank P. Hewett for making the LBQS spectra available for this study, and for helpful responses to our questions.  Most of the data analysis for this project was performed using the ISIS platform \citep{hd00}.  We thank the referee for helpful comments that have improved this study.

We gratefully acknowledge support from NASA LTSA grant NAG5-13035 (RRG, WNB, DPS) and NSF grant AST0607634 (DPS).

Funding for the SDSS and SDSS-II has been provided by the Alfred P. Sloan Foundation, the Participating Institutions, the National Science Foundation, the U.S. Department of Energy, the National Aeronautics and Space Administration, the Japanese Monbukagakusho, the Max Planck Society, and the Higher Education Funding Council for England.  The SDSS Web site \hbox{is {\tt http://www.sdss.org/}.}

The Hobby-Eberly Telescope (HET) is a joint project of the University of Texas at Austin, the Pennsylvania State University,  Stanford University, Ludwig-Maximillians-Universit\"at M\"unchen, and Georg-August-Universit\"at G\"ottingen.  The HET is named in honor of its principal benefactors, William P. Hobby and Robert E. Eberly.  The Marcario Low-Resolution Spectrograph is named for Mike Marcario of High Lonesome Optics, who fabricated several optics for the instrument but died before its completion; it is a joint project of the Hobby-Eberly Telescope partnership and the Instituto de Astronom\'{\i}a de la Universidad Nacional Aut\'onoma de M\'exico.

%% Appendix material should be preceded with a single \appendix command.
%% There should be a \section command for each appendix. Mark appendix
%% subsections with the same markup you use in the main body of the paper.

%% Each Appendix (indicated with \section) will be lettered A, B, C, etc.
%% The equation counter will reset when it encounters the \appendix
%% command and will number appendix equations (A1), (A2), etc.

\appendix{\label{indivSrcSec}}

\section{Notes on Individual Sources}

\begin{figure} [ht]
  \begin{center}
      \includegraphics[width=4in, angle=270]{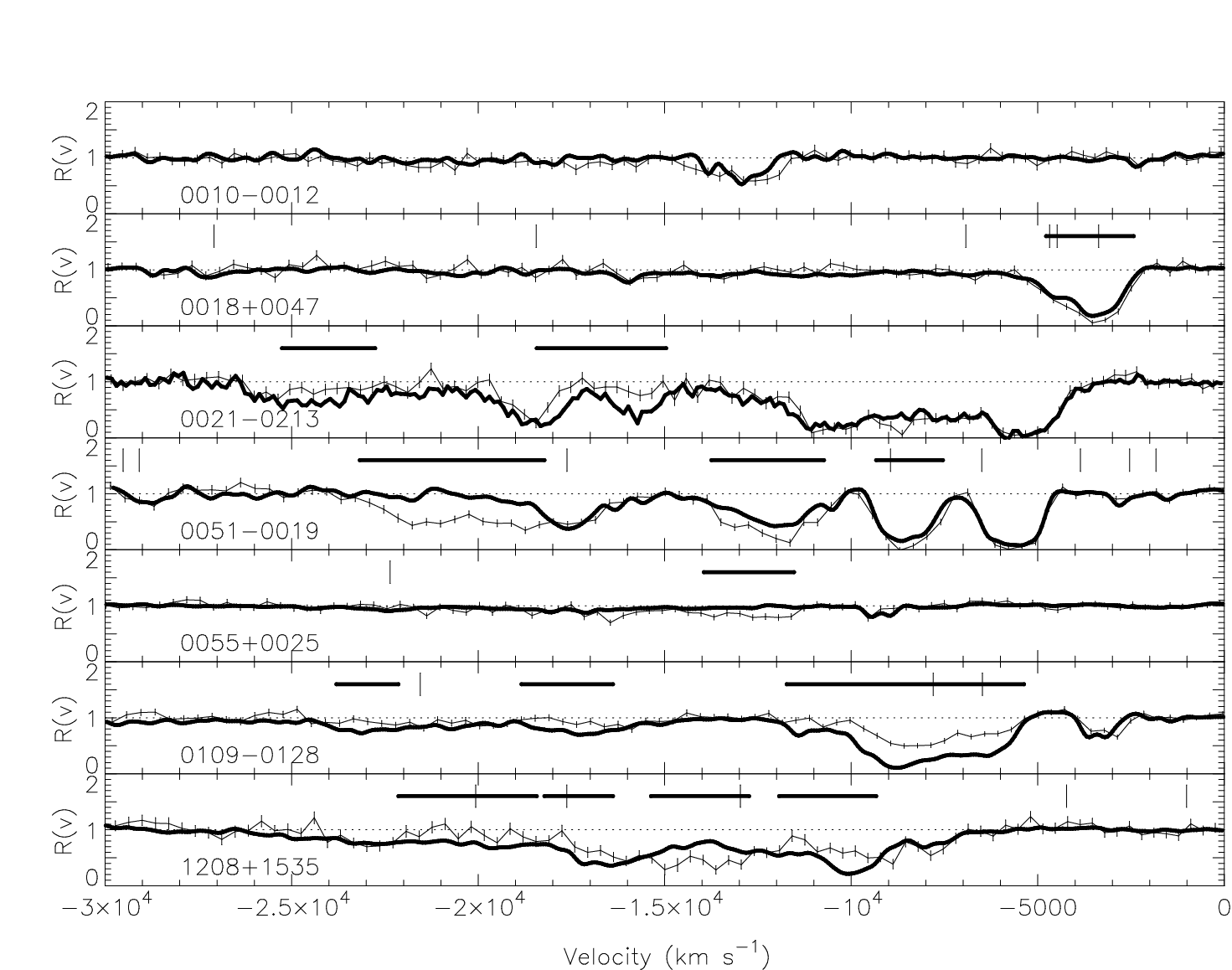}
      \includegraphics[width=4in, angle=270]{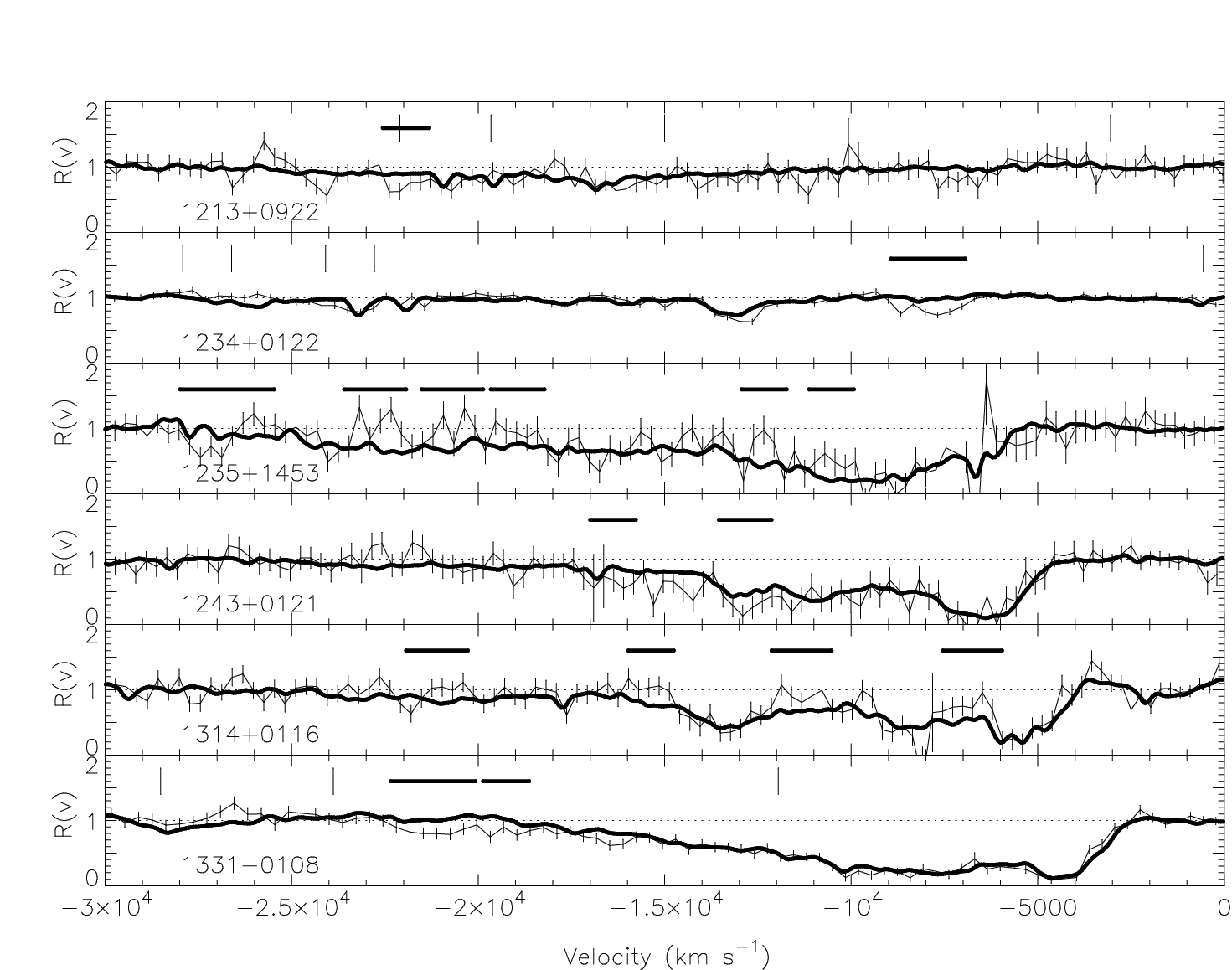}
      \caption{\label{cIVBALsFig}Ratio spectra $R(v)$ (as defined in \S\ref{contFitSec}) of \ion{C}{4} BAL regions for the BAL QSOs in our sample.  The spectra have been divided by the best-fit reddened power law with \ion{C}{4} and \ion{Si}{4} emission lines for each epoch.  LBQS spectra are shown as thin black lines (with error bars), while the SDSS and {\it HET} spectra are shown as thick lines.  The SDSS spectra have been convolved with a Gaussian to approximate the LBQS resolution.  Vertical tick marks above the spectra indicate locations where lines from intervening systems may be present (see \S\ref{balSampleSelectionSec}).  The thick, horizontal lines above the spectra mark regions at least 1200~km~s$^{-1}$ wide which varied between epochs, as described in \S\ref{charVardVSec}.}
   \end{center}
\end{figure}

In this appendix, we briefly discuss the spectra of the \ion{C}{4} regions for the 13 BAL QSOs in our study.  In order to compare \ion{C}{4} absorption spectra from two epochs directly, we constructed the ratio spectrum, $R(\lambda)$, for each epoch by dividing out the continuum and broad line emission as described in \S\ref{contFitSec}.  We convolved the SDSS spectrum with a Gaussian of the appropriate width to approximate the LBQS spectral resolution.  The ratio spectra for the BAL QSOs in our sample are shown in Figure~\ref{cIVBALsFig}.

We have visually searched the SDSS spectra for absorption from narrow doublets up to $\approx100,000$~km~s$^{-1}$ blueward of the broad \ion{Mg}{2} $\lambda$2799 emission line.  These doublets are likely due to \ion{Mg}{2} ions in intervening absorbers.  After determining the velocity offset of these (likely) intervening systems, we search a theoretical line list \citep{vvf96} for lines which, at that velocity, may also appear in the \ion{C}{4} absorption region.  We do not know the ionization state or elemental abundances of the intervening systems, so we cannot predict the strengths of these lines.  However, their velocity widths and equivalent widths are small compared to intrinsic BAL absorption features, and most of these lines do not reside in variable regions anyway (Figure~\ref{cIVBALsFig}), so they will not significantly influence our results.  The locations of (potential) interesting lines with absorption oscillator strengths $> 0.1$ are marked in Figure~\ref{cIVBALsFig}.  The velocities of these intervening absorbers are given in Table~\ref{bALDataTab}.

\subsection{0010--0012\label{0010--0012AnalysisSec}}

A relatively narrow BAL has decreased in width over time.  $BI = 0$ in the SDSS observation.

\subsection{0018+0047\label{0018+0047AnalysisSec}}

This source has a strong BAL at a relatively low outflow velocity (--2000 to \mbox{--6000~km~s$^{-1}$}).  The absorption has perhaps weakened between epochs.  Intervening absorption may be weakly present in the \ion{C}{4} absorption trough.

\subsection{0021--0213\label{0021--0213AnalysisSec}}

In this case, the second epoch is an {\it HET} LRS spectrum, rather than an SDSS spectrum.  Weak variation is seen in narrow components, while a stronger trough has appeared at --16,000~km~s$^{-1}$.

\subsection{0051--0019\label{0051--0019AnalysisSec}}

Four strong, broad \ion{C}{4} absorption troughs are visible at velocities ranging from --5000 to --22,000~km~s$^{-1}$.  The weaker \ion{C}{4} absorption components show the greatest variation.  The low-velocity region of the highest-velocity absorption component (at --18,000~km~s$^{-1}$) has remained nearly constant while the remainder of the high-velocity component has disappeared.  The broad absorption at $\approx$--18,000~km~s$^{-1}$ may, in principle, be contaminated by a narrow line from \ion{Al}{1} in an intervening system, but the full resolution of the SDSS spectrum shows that the absorption at that velocity is too broad ($\approx$2000~km~s$^{-1}$ wide) for the contamination to have any significant effect.

\subsection{0055+0025\label{0055+0025AnalysisSec}}

This source is technically classified as a BAL QSO based on weak, broad \ion{C}{4} absorption in the --11,000 to \hbox{--14,000}~km~s$^{-1}$ range.  The absorption feature is weak, and disappeared between epochs.

\subsection{0109--0128\label{0109--0128AnalysisSec}}

This source has varied dramatically between epochs.  Two broad, high-velocity components have appeared at \hbox{--23,000} and --17,000~km~s$^{-1}$.  Weak absorption features may be present at lower velocities in the LBQS epoch.  The broad, low-velocity BAL between --12,000 and --5,000~km~s$^{-1}$ has strengthened considerably, almost to black at the deepest point.  Between --10,000 and --6,000~km~s$^{-1}$, the difference between ratio spectra is nearly constant, consistent with a scenario in which the covering factor of the source varies for a saturated absorber.  The low-velocity absorption component at \hbox{--3,000}~km~s$^{-1}$ shows evidence of acceleration between epochs, but the putative acceleration (of magnitude $\approx$200~km~s$^{-1}$) is smaller than the accuracy of the wavelength calibration of the LBQS spectrum.

\subsection{1208+1535\label{1208+1535AnalysisSec}}

Dramatic variation is evident in this case.  The deepest points in the \ion{C}{4} absorption and the local maximum between them have shifted several thousand km~s$^{-1}$ to higher velocities.  The onset velocity (about --7000~km~s$^{-1}$) has not changed.  Broad, shallow absorption has increased from --17,000 to --23,000~km~s$^{-1}$.

\subsection{1213+0922\label{1213+0922AnalysisSec}}

Technically classified as a BAL QSO based on our $BI$ measurements, 1213+0922 was not classified as an HF03 BAL QSO.  The absorption is shallow and broad.  We note that similar ratio spectrum shapes are seen in several SDSS QSOs (not observed in the LBQS and therefore not in our sample).  We cannot rule out the possibility that the feature is due to structure in the continuum emission, rather than broad absorption.  (See also \S\ref{identifyBALsSec}.)

\subsection{1234+0122\label{1234+0122AnalysisSec}}

This source was classified as an HF03 BAL QSO, but our continuum fit technically gives $BI = 0$ in both epochs.  Nonetheless, there are broad ($\approx$2000~km~s$^{-1}$ wide), shallow absorption features at --23,000, --13,000, and --8000~km~s$^{-1}$ in the \ion{C}{4} absorption region.  Intervening systems at --79,100 and --75,500~km~s$^{-1}$ (offset from the system velocity) may be responsible for the ``high-velocity'' absorption, although we note in that case that the intervening absorber seems to have varied between epochs.  The variation may also be due to intrinsic BAL components at high velocities.

The broader absorption features in the \ion{C}{4} region at --13,000 and --8000~km~s$^{-1}$ are likely intrinsic.  The higher-velocity component has weakened between epochs, while the lower-velocity component has disappeared altogether.

\subsection{1235+1453\label{1235+1453AnalysisSec}}

Our power law fit appears to match the continuum well in regions free of absorption and emission, so that the differences between epochs appear to be real changes in the absorption.  The broad \ion{C}{4} absorber has deepened between epochs to extend out to --25,000~km~s$^{-1}$.  The absorber transmission is relatively constant from --13,000 out to --25,000~km~s$^{-1}$.

\subsection{1243+0121\label{1243+0121AnalysisSec}}

Similarly to 1235+1453, this source shows broad \ion{C}{4} absorption with a sharp onset at --5000~km~s$^{-1}$.

\subsection{1314+0116\label{1314+0116AnalysisSec}}

This source is similar to 1235+1453 and 1243+0121 in that it shows deep, broad \ion{C}{4} absorption beginning at a detachment velocity of --4000~km~s$^{-1}$.

\subsection{1331--0108\label{1331--0108AnalysisSec}}

\ion{C}{4} absorption from --15,000 to --23,000~km~s$^{-1}$ has apparently weakened gradually between epochs.  1331--0108 is a LoBAL, with broad \ion{Al}{3} and \ion{Mg}{2} absorption evident in the spectrum (not shown).

% -------------- BEGIN BIBLIOGRAPHY -----------------

%\bibliographystyle{apj}
\bibliographystyle{apj3}
\bibliography{apj-jour,bibliography}

% -------------- BEGIN TABLES -----------------

\begin{deluxetable}{lrrllrrr}
\tabletypesize{\scriptsize}
\tablecolumns{8}
\tablewidth{0pc}
\tablecaption{Observation Log\label{obsCatTab}}
\tablehead{\colhead{LBQS} & \colhead{SDSS} & \colhead{$z$} & \colhead{LBQS} & \colhead{SDSS/{\it HET}} & \colhead{$B_J$\tablenotemark{a}} & \colhead{$M_i$\tablenotemark{b}} & \colhead{$\Delta t_{sys}$\tablenotemark{c}} \\ \colhead{B1950} & \colhead{J2000} & \colhead{} & \colhead{Date} & \colhead{Date} & \colhead{} & \colhead{} & \colhead{(yr)}}
\startdata
\input{tab1.tex}
\enddata
\tablenotetext{a}{The $B_J$ magnitude given in the LBQS QSO catalog \citep{hfc95}.}
\tablenotetext{b}{The absolute $i$ magnitude given in the SDSS DR5 QSO catalog \citep{s+07}.}
\tablenotetext{c}{The rest-frame time in years between LBQS and SDSS observations.}
\end{deluxetable}

\begin{deluxetable}{lrrrrrrr}
\tabletypesize{\scriptsize}
\tablecolumns{8}
\tablewidth{0pc}
\tablecaption{Continuum and \ion{C}{4} BAL Data\label{bALDataTab}}
\tablehead{\colhead{LBQS} & \colhead{LBQS} & \colhead{SDSS} & \colhead{LBQS} & \colhead{SDSS} & \colhead{LBQS \ion{C}{4}} & \colhead{SDSS/{\it HET} \ion{C}{4}} & \colhead{Intervening System} \\ \colhead{B1950} & \colhead{$L_{\nu}(1400 $\AA$)$\tablenotemark{a}} & \colhead{$L_{\nu}(1400 $\AA$)$\tablenotemark{a}} & \colhead{$L_{\nu}(2500 $\AA$)$\tablenotemark{a}} & \colhead{$L_{\nu}(2500 $\AA$)$\tablenotemark{a}} & \colhead{EW (\AA)\tablenotemark{b}} & \colhead{EW (\AA)\tablenotemark{b}} & \colhead{Velocities\tablenotemark{c}}}
\startdata
\input{tab2.tex}
\enddata
\tablenotetext{a}{Monochromatic luminosities are given in units of 10$^{31}$ erg s$^{-1}$ Hz$^{-1}$ and include a factor of $(1+z)$ ``bandpass correction'' \citep{h00}.  0021--0213 was not observed with the SDSS, and we do not include it in our continuum variation studies.}
\tablenotetext{b}{In the \ion{C}{4} region, 1~\AA\ is $\approx$200~km~s$^{-1}$.}
\tablenotetext{c}{Velocities of any intervening systems are given in units of 1000~km~s$^{-1}$ offset from the rest frame.  The corresponding redshift for each system is given in parentheses.}
\end{deluxetable}

\begin{deluxetable}{lcrrrr}
\tabletypesize{\scriptsize}
\tablecolumns{6}
\tablewidth{0pc}
\tablecaption{C{\sc IV} Balnicity Indices\label{bITab}\tablenotemark{a}}
\tablehead{\colhead{LBQS B1950} & \colhead{HF03} & \colhead{LBQS $BI$} & \colhead{SDSS/{\it HET} $BI$} & \colhead{LBQS $BI_0$} & \colhead{SDSS/{\it HET} $BI_0$} \\ \colhead{} & \colhead{BAL?} & \colhead{(km s$^{-1}$)} & \colhead{(km s$^{-1}$)} & \colhead{(km s$^{-1}$)} & \colhead{(km s$^{-1}$)}}
\startdata
\input{tab3.tex}
\enddata
\tablenotetext{a}{$BI$ and $BI_0$ were calculated for the LBQS, SDSS, and HET spectra using our continuum fits as described in \S\ref{identifyBALsSec}.}
\end{deluxetable}

\begin{deluxetable}{rrrr}
\tabletypesize{\scriptsize}
\tablecolumns{4}
\tablewidth{0pc}
\tablecaption{Varying Regions\label{varRegTab}}
\tablehead{\colhead{LBQS} & \colhead{$\Sigma_{\Delta v}$\tablenotemark{a}} & \colhead{$\Delta v$} & \colhead{$\langle v \rangle$} \\ \colhead{B1950} & \colhead{(km s$^{-1}$)} & \colhead{(km s$^{-1}$)} & \colhead{(km s$^{-1}$)}}
\startdata
\input{tab4.tex}
\enddata
\tablenotetext{a}{$\Sigma_{\Delta v}$ is the sum of velocity widths of all varying regions in a source, so only one value is given per source.  For sources not listed in this table, we found no varying regions, and $\Sigma_{\Delta v} = 0$.}
\end{deluxetable}

\begin{deluxetable}{rr}
\tabletypesize{\scriptsize}
\tablecolumns{2}
\tablewidth{0pc}
\tablecaption{Trough Onset Acceleration Limits\label{accelLimitTab}}
\tablehead{\colhead{LBQS} & \colhead{Upper Limit} \\ \colhead{B1950} & \colhead{(cm s$^{-2}$)}}
\startdata
\input{tab5.tex}
\enddata
\end{deluxetable}

% -------------- BEGIN FIGURES -----------------

\clearpage

\clearpage

\clearpage

\clearpage

\clearpage

\clearpage

\end{document}

%% file: tab1.tex
$0010-0012$ & $ 001306.15+000431.9 $ & 2.16 & 1987 Sep 25 & 2000 Sep 6 & 18.5 & $-27.1$ & 4.1\\
$0018+0047$ & $ 002127.88+010420.2 $ & 1.82 & 1988 Nov 2 & 2000 Dec 7 & 17.8 & $-26.8$ & 4.3\\
$0021-0213$ & \dots & 2.29 & 1988 Sep 9 & 2001 Dec 10 & 18.7 & $\dots$ & 4.0\\
$0051-0019$ & $ 005355.15-000309.3 $ & 1.72 & 1989 Sep 29 & 2000 Aug 27 & 18.7 & $-27.1$ & 4.0\\
$0055+0025$ & $ 005824.75+004113.3 $ & 1.92 & 1988 Sep 9 & 2000 Aug 27 & 17.5 & $-28.6$ & 4.1\\
$0109-0128$ & $ 011227.60-011221.7 $ & 1.76 & 1989 Sep 29 & 2000 Sep 7 & 18.3 & $-27.3$ & 4.0\\
$1208+1535$ & $ 121125.49+151851.5 $ & 1.96 & 1987 Mar 26 & 2005 Apr 6 & 17.9 & $-27.4$ & 6.1\\
$1213+0922$ & $ 121539.66+090607.4 $ & 2.72 & 1987 Mar 27 & 2003 Feb 2 & 18.1 & $-28.0$ & 4.3\\
$1234+0122$ & $ 123724.51+010615.4 $ & 2.02 & 1987 Apr 2 & 2001 Feb 1 & 18.0 & $-27.7$ & 4.6\\
$1235+1453$ & $ 123736.42+143640.2 $ & 2.68 & 1986 Dec 29 & 2005 Mar 13 & 18.6 & $-27.7$ & 4.9\\
$1243+0121$ & $ 124551.44+010505.0 $ & 2.81 & 1987 Apr 2 & 2001 Jan 19 & 18.5 & $-28.3$ & 3.6\\
$1314+0116$ & $ 131714.21+010013.0 $ & 2.70 & 1988 Mar 16 & 2001 Mar 16 & 18.6 & $-28.0$ & 3.5\\
$1331-0108$ & $ 133428.06-012349.0 $ & 1.88 & 1988 Mar 21 & 2002 Jun 1 & 17.9 & $-28.4$ & 4.9\\

%% file: tab2.tex
$0010-0012$ & $0.5$ & $1.4$ & $0.3$ & $1.6$ & $-7.3 \pm 1.3$ & $-3.8 \pm 0.8$ & \dots\\
$0018+0047$ & $1.4$ & $0.8$ & $1.2$ & $1.1$ & $-11.2 \pm 1.2$ & $-13.5 \pm 0.5$ & --27 (1.58), --57 (1.33)\\
$0021-0213$ & $1.8$ & $\dots$ & $2.5$ & $\dots$ & $-37.6 \pm 1.4$ & $-51.0 \pm 0.8$ & \dots\\
$0051-0019$ & $1.2$ & $0.5$ & $2.3$ & $1.5$ & $-40.6 \pm 1.1$ & $-29.1 \pm 0.8$ & --29 (1.47), --56.2 (1.25)\\
$0055+0025$ & $4.7$ & $4.2$ & $5.1$ & $6.1$ & $-6.3 \pm 1.0$ & $-4.0 \pm 0.4$ & --99 (1.07)\\
$0109-0128$ & $4.3$ & $1.2$ & $4.1$ & $1.9$ & $-12.6 \pm 1.0$ & $-29.5 \pm 0.6$ & --60 (1.25)\\
$1208+1535$ & $1.5$ & $0.8$ & $1.1$ & $1.5$ & $-25.5 \pm 1.5$ & $-31.9 \pm 0.5$ & --40 (1.59), --82.5 (1.23)\\
$1213+0922$ & $2.5$ & $2.3$ & $3.2$ & $3.8$ & $-13.8 \pm 2.0$ & $-9.5 \pm 0.5$ & --42 (2.23)\\
$1234+0122$ & $1.3$ & $1.8$ & $1.6$ & $2.7$ & $-6.4 \pm 0.9$ & $-5.7 \pm 0.5$ & --75.5 (1.33), --79.1 (1.30)\\
$1235+1453$ & $4.1$ & $1.0$ & $4.6$ & $1.7$ & $-29.2 \pm 2.7$ & $-41.0 \pm 0.7$ & \dots\\
$1243+0121$ & $2.4$ & $3.0$ & $3.2$ & $4.7$ & $-34.4 \pm 2.5$ & $-34.0 \pm 0.5$ & \dots\\
$1314+0116$ & $2.0$ & $2.4$ & $1.5$ & $3.9$ & $-22.5 \pm 1.9$ & $-30.9 \pm 0.5$ & \dots\\
$1331-0108$ & $0.9$ & $0.9$ & $2.3$ & $4.9$ & $-44.8 \pm 1.2$ & $-41.8 \pm 0.5$ & --50.7 (1.43)\\

%% file: tab3.tex
$0010-0012$ & No & 48 & 0 & 48 & 0\\
$0018+0047$ & Yes & 40 & 39 & 309 & 257\\
$0021-0213$ & Yes & 4042 & 6108 & 4042 & 6108\\
$0051-0019$ & Yes & 2474 & 191 & 2474 & 191\\
$0055+0025$ & No & 61 & 0 & 61 & 0\\
$0109-0128$ & Yes & 746 & 1939 & 746 & 1939\\
$1208+1535$ & Yes & 2517 & 3302 & 2517 & 3302\\
$1213+0922$ & No & 47 & 25 & 47 & 25\\
$1234+0122$ & Yes & 0 & 0 & 0 & 0\\
$1235+1453$ & Yes & 1342 & 4820 & 1342 & 4820\\
$1243+0121$ & Yes & 3622 & 3764 & 3622 & 3764\\
$1314+0116$ & Yes & 257 & 3077 & 257 & 3077\\
$1331-0108$ & Yes & 6568 & 5977 & 6568 & 6199\\

%% file: tab4.tex
$0018+0047$ & $2300$ & $  2300$ & $ -3600$\\
&&&\\
$0021-0213$ & $6000$ & $  2500$ & $-24000$\\
$$ & $$ & $  3500$ & $-16700$\\
&&&\\
$0051-0019$ & $9800$ & $  5000$ & $-20700$\\
$$ & $$ & $  3000$ & $-12200$\\
$$ & $$ & $  1800$ & $ -8400$\\
&&&\\
$0055+0025$ & $2400$ & $  2400$ & $-12700$\\
&&&\\
$0109-0128$ & $10500$ & $  1700$ & $-23000$\\
$$ & $$ & $  2500$ & $-17600$\\
$$ & $$ & $  6400$ & $ -8600$\\
&&&\\
$1208+1535$ & $10800$ & $  3700$ & $-20300$\\
$$ & $$ & $  1800$ & $-17300$\\
$$ & $$ & $  2600$ & $-14100$\\
$$ & $$ & $  2600$ & $-10600$\\
&&&\\
$1213+0922$ & $1200$ & $  1200$ & $-21900$\\
&&&\\
$1234+0122$ & $2000$ & $  2000$ & $ -7900$\\
&&&\\
$1235+1453$ & $9700$ & $  2500$ & $-26700$\\
$$ & $$ & $  1700$ & $-22800$\\
$$ & $$ & $  1700$ & $-20700$\\
$$ & $$ & $  1400$ & $-18900$\\
$$ & $$ & $  1200$ & $-12300$\\
$$ & $$ & $  1200$ & $-10500$\\
&&&\\
$1243+0121$ & $2600$ & $  1200$ & $-16400$\\
$$ & $$ & $  1400$ & $-12800$\\
&&&\\
$1314+0116$ & $6100$ & $  1700$ & $-21100$\\
$$ & $$ & $  1200$ & $-15400$\\
$$ & $$ & $  1600$ & $-11300$\\
$$ & $$ & $  1600$ & $ -6800$\\
&&&\\
$1331-0108$ & $3500$ & $  2300$ & $-21200$\\
$$ & $$ & $  1200$ & $-19300$\\

%% file: tab5.tex
$0018+0047$ & $0.14$\\
$0021-0213$ & $0.15$\\
$0051-0019$ & $0.15$\\
$1235+1453$ & $0.12$\\
$1243+0121$ & $0.17$\\
$1314+0116$ & $0.17$\\
$1331-0108$ & $0.12$\\

%% file: ms.bbl
\begin{thebibliography}{69}
\expandafter\ifx\csname natexlab\endcsname\relax\def\natexlab#1{#1}\fi
\expandafter\ifx\csname url\endcsname\relax
  \def\url#1{{\tt #1}}\fi
\expandafter\ifx\csname urlprefix\endcsname\relax\def\urlprefix{URL }\fi
\providecommand{\eprint}[2][]{\url{#2}}

\bibitem[\protect\astroncite{{Arav} et~al.}{1999{\natexlab{a}}}]{ablgwbd99}
{Arav}, N., {Becker}, R.~H., {Laurent-Muehleisen}, S.~A., {Gregg}, M.~D.,
  {White}, R.~L., {Brotherton}, M.~S., \& {de Kool}, M. 1999{\natexlab{a}},
  \apj, 524, 566

\bibitem[\protect\astroncite{{Arav} et~al.}{1999{\natexlab{b}}}]{akdjb99}
{Arav}, N., {Korista}, K.~T., {de Kool}, M., {Junkkarinen}, V.~T., \&
  {Begelman}, M.~C. 1999{\natexlab{b}}, \apj, 516, 27

\bibitem[\protect\astroncite{{Arav} et~al.}{2001}]{a+01}
{Arav}, N., et~al. 2001, \apj, 561, 118

\bibitem[\protect\astroncite{{Barlow}}{1993}]{b93}
{Barlow}, T.~A. 1993, Ph.D. thesis, AA(California Univ.)

\bibitem[\protect\astroncite{{Barlow} et~al.}{1989}]{bjb89}
{Barlow}, T.~A., {Junkkarinen}, V.~T., \& {Burbidge}, E.~M. 1989, \apj, 347,
  674

\bibitem[\protect\astroncite{{Barlow} et~al.}{1992}]{bjbwmk92}
{Barlow}, T.~A., {Junkkarinen}, V.~T., {Burbidge}, E.~M., {Weymann}, R.~J.,
  {Morris}, S.~L., \& {Korista}, K.~T. 1992, \apj, 397, 81

\bibitem[\protect\astroncite{{Becker} et~al.}{2000}]{bwgbla00}
{Becker}, R.~H., {White}, R.~L., {Gregg}, M.~D., {Brotherton}, M.~S.,
  {Laurent-Muehleisen}, S.~A., \& {Arav}, N. 2000, \apj, 538, 72

\bibitem[\protect\astroncite{{Boksenberg} et~al.}{1977}]{bcafps77}
{Boksenberg}, A., {Carswell}, R.~F., {Allen}, D.~A., {Fosbury}, R.~A.~E.,
  {Penston}, M.~V., \& {Sargent}, W.~L.~W. 1977, \mnras, 178, 451

\bibitem[\protect\astroncite{{Boroson} et~al.}{1991}]{bmmp91}
{Boroson}, T.~A., {Meyers}, K.~A., {Morris}, S.~L., \& {Persson}, S.~E. 1991,
  \apjl, 370, L19

\bibitem[\protect\astroncite{{Brotherton} et~al.}{2001}]{btbglw01}
{Brotherton}, M.~S., {Tran}, H.~D., {Becker}, R.~H., {Gregg}, M.~D.,
  {Laurent-Muehleisen}, S.~A., \& {White}, R.~L. 2001, \apj, 546, 775,
  \eprint{arXiv:astro-ph/0008396}

\bibitem[\protect\astroncite{{Canalizo} \& {Stockton}}{2001}]{cs01}
{Canalizo}, G. \& {Stockton}, A. 2001, \apj, 555, 719

\bibitem[\protect\astroncite{{Cardelli} et~al.}{1989}]{ccm89}
{Cardelli}, J.~A., {Clayton}, G.~C., \& {Mathis}, J.~S. 1989, \apj, 345, 245

\bibitem[\protect\astroncite{{de Kool} et~al.}{2001}]{dabgwlpk01}
{de Kool}, M., {Arav}, N., {Becker}, R.~H., {Gregg}, M.~D., {White}, R.~L.,
  {Laurent-Muehleisen}, S.~A., {Price}, T., \& {Korista}, K.~T. 2001, \apj,
  548, 609

\bibitem[\protect\astroncite{{Desroches} et~al.}{2006}]{dfklmglmsbhnop06}
{Desroches}, L.-B., et~al. 2006, \apj, 650, 88

\bibitem[\protect\astroncite{{Foltz} et~al.}{1987{\natexlab{a}}}]{fchmtwa87}
{Foltz}, C.~B., {Chaffee}, F.~H., Jr., {Hewett}, P.~C., {MacAlpine}, G.~M.,
  {Turnshek}, D.~A., {Weymann}, R.~J., \& {Anderson}, S.~F. 1987{\natexlab{a}},
  \aj, 94, 1423

\bibitem[\protect\astroncite{{Foltz} et~al.}{1987{\natexlab{b}}}]{fwmt87}
{Foltz}, C.~B., {Weymann}, R.~J., {Morris}, S.~L., \& {Turnshek}, D.~A.
  1987{\natexlab{b}}, \apj, 317, 450

\bibitem[\protect\astroncite{{Gallagher} et~al.}{2006}]{gbcpgs06}
{Gallagher}, S.~C., {Brandt}, W.~N., {Chartas}, G., {Priddey}, R., {Garmire},
  G.~P., \& {Sambruna}, R.~M. 2006, \apj, 644, 709

\bibitem[\protect\astroncite{{Gallagher} et~al.}{2004}]{gbwccl04}
{Gallagher}, S.~C., {Brandt}, W.~N., {Wills}, B.~J., {Charlton}, J.~C.,
  {Chartas}, G., \& {Laor}, A. 2004, \apj, 603, 425

\bibitem[\protect\astroncite{{Gallagher} et~al.}{2005}]{gssbchhb05}
{Gallagher}, S.~C., {Schmidt}, G.~D., {Smith}, P.~S., {Brandt}, W.~N.,
  {Chartas}, G., {Hylton}, S., {Hines}, D.~C., \& {Brotherton}, M.~S. 2005,
  \apj, 633, 71

\bibitem[\protect\astroncite{{Ganguly} et~al.}{2001}]{gce01}
{Ganguly}, R., {Charlton}, J.~C., \& {Eracleous}, M. 2001, \apjl, 556, L7

\bibitem[\protect\astroncite{{Gregg} et~al.}{2006}]{gbd06}
{Gregg}, M.~D., {Becker}, R.~H., \& {de Vries}, W. 2006, \apj, 641, 210

\bibitem[\protect\astroncite{{Gregg} et~al.}{2002}]{gbwrcf02}
{Gregg}, M.~D., {Becker}, R.~H., {White}, R.~L., {Richards}, G.~T., {Chaffee},
  F.~H., \& {Fan}, X. 2002, \apjl, 573, L85

\bibitem[\protect\astroncite{{Hall} et~al.}{2007}]{hsher07}
{Hall}, P.~B., {Sadavoy}, S.~I., {Hutsemekers}, D., {Everett}, J.~E., \&
  {Rafiee}, A. 2007, ArXiv e-prints, 704

\bibitem[\protect\astroncite{{Hall} et~al.}{2002}]{h+02}
{Hall}, P.~B., et~al. 2002, \apjs, 141, 267

\bibitem[\protect\astroncite{{Hamann} et~al.}{1995}]{hbbbcjl95}
{Hamann}, F., {Barlow}, T.~A., {Beaver}, E.~A., {Burbidge}, E.~M., {Cohen},
  R.~D., {Junkkarinen}, V., \& {Lyons}, R. 1995, \apj, 443, 606

\bibitem[\protect\astroncite{{Hamann} et~al.}{1997}]{hbj97}
{Hamann}, F., {Barlow}, T.~A., \& {Junkkarinen}, V. 1997, \apj, 478, 87

\bibitem[\protect\astroncite{{Hewett} \& {Foltz}}{2003}]{hf03}
{Hewett}, P.~C. \& {Foltz}, C.~B. 2003, \aj, 125, 1784

\bibitem[\protect\astroncite{{Hewett} et~al.}{1995}]{hfc95}
{Hewett}, P.~C., {Foltz}, C.~B., \& {Chaffee}, F.~H. 1995, \aj, 109, 1498

\bibitem[\protect\astroncite{{Hewett} et~al.}{1991}]{hfcfwmam91}
{Hewett}, P.~C., {Foltz}, C.~B., {Chaffee}, F.~H., {Francis}, P.~J., {Weymann},
  R.~J., {Morris}, S.~L., {Anderson}, S.~F., \& {MacAlpine}, G.~M. 1991, \aj,
  101, 1121

\bibitem[\protect\astroncite{{Hogg}}{1999}]{h00}
{Hogg}, D.~W. 1999, ArXiv Astrophysics e-prints, astro-ph/9905116,
  \eprint{astro-ph/9905116}

\bibitem[\protect\astroncite{{Hopkins} et~al.}{2004}]{hshrcsvjbs04}
{Hopkins}, P.~F., et~al. 2004, \aj, 128, 1112

\bibitem[\protect\astroncite{{Houck} \& {Denicola}}{2000}]{hd00}
{Houck}, J.~C. \& {Denicola}, L.~A. 2000, in ASP Conf. Ser. 216: Astronomical
  Data Analysis Software and Systems IX, eds. N.~{Manset}, C.~{Veillet}, \&
  D.~{Crabtree}, 591--+

\bibitem[\protect\astroncite{{Junkkarinen} et~al.}{2001}]{jsbbchl01}
{Junkkarinen}, V., {Shields}, G.~A., {Beaver}, E.~A., {Burbidge}, E.~M.,
  {Cohen}, R.~D., {Hamann}, F., \& {Lyons}, R.~W. 2001, \apjl, 549, L155

\bibitem[\protect\astroncite{{Kaspi} et~al.}{2004}]{kbcer04}
{Kaspi}, S., {Brandt}, W.~N., {Collinge}, M.~J., {Elvis}, M., \& {Reynolds},
  C.~S. 2004, \aj, 127, 2631

\bibitem[\protect\astroncite{{Kaspi} et~al.}{2007}]{kbmnss07}
{Kaspi}, S., {Brandt}, W.~N., {Maoz}, D., {Netzer}, H., {Schneider}, D.~P., \&
  {Shemmer}, O. 2007, \apj, 659, 997, \eprint{arXiv:astro-ph/0612722}

\bibitem[\protect\astroncite{{Kaspi} et~al.}{2000}]{ksnmjg00}
{Kaspi}, S., {Smith}, P.~S., {Netzer}, H., {Maoz}, D., {Jannuzi}, B.~T., \&
  {Giveon}, U. 2000, \apj, 533, 631

\bibitem[\protect\astroncite{{Korista} et~al.}{1993}]{kvmw93}
{Korista}, K.~T., {Voit}, G.~M., {Morris}, S.~L., \& {Weymann}, R.~J. 1993,
  \apjs, 88, 357

\bibitem[\protect\astroncite{{Krolik} \& {Kriss}}{2001}]{kk01}
{Krolik}, J.~H. \& {Kriss}, G.~A. 2001, \apj, 561, 684

\bibitem[\protect\astroncite{{Lundgren} et~al.}{2007}]{lwbhsyvb07}
{Lundgren}, B.~F., {Wilhite}, B.~C., {Brunner}, R.~J., {Hall}, P.~B.,
  {Schneider}, D.~P., {York}, D.~G., {Vanden Berk}, D.~E., \& {Brinkmann}, J.
  2007, \apj, 656, 73

\bibitem[\protect\astroncite{{Ma}}{2002}]{m02}
{Ma}, F. 2002, \mnras, 335, L99

\bibitem[\protect\astroncite{{Maoz} et~al.}{2002}]{mmen02}
{Maoz}, D., {Markowitz}, A., {Edelson}, R., \& {Nandra}, K. 2002, \aj, 124,
  1988

\bibitem[\protect\astroncite{{Michalitsianos} et~al.}{1996}]{mon96}
{Michalitsianos}, A.~G., {Oliversen}, R.~J., \& {Nichols}, J. 1996, \apj, 461,
  593

\bibitem[\protect\astroncite{{Miller} et~al.}{2006}]{mbglwgs06}
{Miller}, B.~P., {Brandt}, W.~N., {Gallagher}, S.~C., {Laor}, A., {Wills},
  B.~J., {Garmire}, G.~P., \& {Schneider}, D.~P. 2006, \apj, 652, 163

\bibitem[\protect\astroncite{{Misawa} et~al.}{2007}]{meck07}
{Misawa}, T., {Eracleous}, M., {Charlton}, J.~C., \& {Kashikawa}, N. 2007,
  \apj, 660, 152

\bibitem[\protect\astroncite{{Morris} et~al.}{1991}]{mwahffcm91}
{Morris}, S.~L., {Weymann}, R.~J., {Anderson}, S.~F., {Hewett}, P.~C.,
  {Francis}, P.~J., {Foltz}, C.~B., {Chaffee}, F.~H., \& {MacAlpine}, G.~M.
  1991, \aj, 102, 1627

\bibitem[\protect\astroncite{{Morton}}{1991}]{m91}
{Morton}, D.~C. 1991, \apjs, 77, 119

\bibitem[\protect\astroncite{{Murray} et~al.}{1995}]{mcgv95}
{Murray}, N., {Chiang}, J., {Grossman}, S.~A., \& {Voit}, G.~M. 1995, \apj,
  451, 498

\bibitem[\protect\astroncite{{Narayanan} et~al.}{2004}]{nhbbcjl04}
{Narayanan}, D., {Hamann}, F., {Barlow}, T., {Burbidge}, E.~M., {Cohen}, R.~D.,
  {Junkkarinen}, V., \& {Lyons}, R. 2004, \apj, 601, 715

\bibitem[\protect\astroncite{{O'Donnell}}{1994}]{o94}
{O'Donnell}, J.~E. 1994, \apj, 422, 158

\bibitem[\protect\astroncite{{Ogle} et~al.}{1999}]{ocmtgm99}
{Ogle}, P.~M., {Cohen}, M.~H., {Miller}, J.~S., {Tran}, H.~D., {Goodrich},
  R.~W., \& {Martel}, A.~R. 1999, \apjs, 125, 1

\bibitem[\protect\astroncite{{Pei}}{1992}]{p92}
{Pei}, Y.~C. 1992, \apj, 395, 130

\bibitem[\protect\astroncite{{Press} et~al.}{2002}]{ptvf02}
{Press}, W.~H., {Teukolsky}, S.~A., {Vetterling}, W.~T., \& {Flannery}, B.~P.
  2002, {Numerical Recipes in C++, Second Edition} ({Cambridge University
  Press})

\bibitem[\protect\astroncite{{Proga} et~al.}{2000}]{psk00}
{Proga}, D., {Stone}, J.~M., \& {Kallman}, T.~R. 2000, \apj, 543, 686

\bibitem[\protect\astroncite{{Reichard} et~al.}{2003}]{rrhsvfykb03}
{Reichard}, T.~A., et~al. 2003, \aj, 126, 2594

\bibitem[\protect\astroncite{{Rupke} et~al.}{2002}]{rvs02}
{Rupke}, D.~S., {Veilleux}, S., \& {Sanders}, D.~B. 2002, \apj, 570, 588

\bibitem[\protect\astroncite{{Schlegel} et~al.}{1998}]{sfd98}
{Schlegel}, D.~J., {Finkbeiner}, D.~P., \& {Davis}, M. 1998, \apj, 500, 525

\bibitem[\protect\astroncite{{Schneider} et~al.}{2007}]{s+07}
{Schneider}, D.~P., et~al. 2007, \aj, 134, 102

\bibitem[\protect\astroncite{{Shemmer} et~al.}{2003}]{sunm03}
{Shemmer}, O., {Uttley}, P., {Netzer}, H., \& {McHardy}, I.~M. 2003, \mnras,
  343, 1341

\bibitem[\protect\astroncite{{Smith} \& {Penston}}{1988}]{sp88}
{Smith}, L.~J. \& {Penston}, M.~V. 1988, \mnras, 235, 551

\bibitem[\protect\astroncite{{Trump} et~al.}{2006}]{thrrsvkafbkn06}
{Trump}, J.~R., et~al. 2006, \apjs, 165, 1

\bibitem[\protect\astroncite{{Turnshek} et~al.}{1988}]{tgfw88}
{Turnshek}, D.~A., {Grillmair}, C.~J., {Foltz}, C.~B., \& {Weymann}, R.~J.
  1988, \apj, 325, 651

\bibitem[\protect\astroncite{{Verner} et~al.}{1996}]{vvf96}
{Verner}, D.~A., {Verner}, E.~M., \& {Ferland}, G.~J. 1996, Atomic Data and
  Nuclear Data Tables, 64, 1

\bibitem[\protect\astroncite{{Vilkoviskij} \& {Irwin}}{2001}]{vi01}
{Vilkoviskij}, E.~Y. \& {Irwin}, M.~J. 2001, \mnras, 321, 4

\bibitem[\protect\astroncite{{Voit} et~al.}{1993}]{vwk93}
{Voit}, G.~M., {Weymann}, R.~J., \& {Korista}, K.~T. 1993, \apj, 413, 95

\bibitem[\protect\astroncite{{Weymann} et~al.}{1991}]{wmfh91}
{Weymann}, R.~J., {Morris}, S.~L., {Foltz}, C.~B., \& {Hewett}, P.~C. 1991,
  \apj, 373, 23

\bibitem[\protect\astroncite{{Weymann} et~al.}{1997}]{wmgh97}
{Weymann}, R.~J., {Morris}, S.~L., {Gray}, M.~E., \& {Hutchings}, J.~B. 1997,
  \apj, 483, 717

\bibitem[\protect\astroncite{{Wilhite} et~al.}{2005}]{wvkspbrb05}
{Wilhite}, B.~C., {Vanden Berk}, D.~E., {Kron}, R.~G., {Schneider}, D.~P.,
  {Pereyra}, N., {Brunner}, R.~J., {Richards}, G.~T., \& {Brinkmann}, J.~V.
  2005, \apj, 633, 638

\bibitem[\protect\astroncite{{Wise} et~al.}{2004}]{wecg04}
{Wise}, J.~H., {Eracleous}, M., {Charlton}, J.~C., \& {Ganguly}, R. 2004, \apj,
  613, 129

\bibitem[\protect\astroncite{{York} et~al.}{2000}]{y+00}
{York}, D.~G., et~al. 2000, \aj, 120, 1579

\end{thebibliography}
